\documentclass[fleqn,usenatbib]{mnras}

\usepackage{newtxtext,newtxmath}
\usepackage[percent]{overpic}

\usepackage[T1]{fontenc}

\DeclareRobustCommand{\VAN}[3]{#2}
\let\VANthebibliography\thebibliography
\def\thebibliography{\DeclareRobustCommand{\VAN}[3]{##3}\VANthebibliography}
\newcommand{\nicer}{\textit{NICER}}

\usepackage{graphicx}	
\usepackage{amsmath}	
\usepackage{placeins}

\title[Accretion geometry of MAXI~J1820+070]{Accretion geometry of the black hole binary MAXI J1820+070 probed by frequency-resolved spectroscopy}

\author[M. Axelsson \& A. Veledina]{
Magnus Axelsson$^{1, 2}$\thanks{E-mail: magnusa@astro.su.se, alexandra.veledina@gmail.com} and
Alexandra Veledina$^{3, 4, 5}$
\\
$^{1}$Oskar Klein Center for CosmoParticle Physics, Department of Physics, Stockholm University, SE-10691 Stockholm, Sweden\\ 
$^{2}$Department of Astronomy, Stockholm University, SE-10691 Stockholm, Sweden\\
$^{3}$Department of Physics and Astronomy, FI-20014 University of Turku, Finland\\
$^{4}$Nordita, KTH Royal Institute of Technology and Stockholm University, Roslagstullsbacken 23, SE-10691 Stockholm, Sweden\\
$^{5}$Space Research Institute of the Russian Academy of Sciences, Profsoyuznaya Str. 84/32, 117997 Moscow, Russia
}

\pubyear{2021}

\begin{document}
\label{firstpage}
\pagerange{\pageref{firstpage}--\pageref{lastpage}}
\maketitle

\begin{abstract}
The geometry of the inner accretion flow in the hard and hard-intermediate states of X-ray binaries remains controversial. Using {\nicer} observations of the black hole X-ray binary MAXI~J1820+070 during the rising phase of its 2018 outburst, we study the evolution of the timing properties, in particular the characteristic variability frequencies of the prominent iron K$\alpha$ line. 
Using frequency-resolved spectroscopy, we find that reflection occurs at large distances from the Comptonizing region in the bright hard state. 
During the hard- to soft transition, the variability properties suggest the reflector moves closer to the X-ray source. 
In parallel, the peak of the iron line shifts from 6.5 to $\sim$7\,keV, becoming consistent with that expected of from a highly inclined disc extending close to the black hole.
We additionally find significant changes in the dependence of the root-mean-square (rms) variability on both energy and Fourier frequency as the source softens. 
The evolution of the rms-energy dependence, the line profile, and the timing properties of the iron line as traced by the frequency-resolved spectroscopy all support the picture of a truncated disc/inner flow geometry.
\end{abstract}


\FloatBarrier

\section{Introduction}
\label{introduction}
Black hole X-ray binaries (BH XRBs) display a large range of spectral states, linked to a striking variety in variability properties. Spectral changes are coupled to changes in the geometry of the inner accretion flow, driven by a varying accretion rate. Most sources are transient, and during an outburst progress through a sequence of spectral-timing states on a timescale of weeks to months \citep[e.g.,][]{ZG04}. The states can be broadly classified as ``hard'' or ``soft'' based on the spectral hardness, with more detailed state definitions being made based on parameters such as spectral shape, broad-band variability, and the presence or absence of quasi-periodic oscillations and jets \citep[e.g.,][]{RM06,BM16}.
Variations of the physical parameters and geometry of these components can explain the large differences between the spectral states.  
Spectral studies clearly indicate the presence of a cool, geometrically thin and optically thick accretion disc \citep{SS73} in the soft state; in the low/hard state, a region of hot plasma where soft photons are Comptonized into a hard spectrum has been identified \citep[see, e.g.,][for reviews]{ZG04,RM06,DGK07}.
However, there is currently no consensus on the location of this region.

Hard X-rays emitted by the hot flow may be intercepted by the accretion disc, giving rise to Compton reflection. This reprocessing leads to characteristic features in the spectrum, with one of the most prominent being the Fe K$\alpha$ line at 6.4\,keV \citep{basko74,Fabian89,george91}. 
The iron line complex is often used as a probe of the inner accretion disc; for example, relativistic broadening of the iron line can be used to estimate the inner disc radius, and by extension the spin parameter of the black hole \citep[e.g.,][]{Brenneman06,Miller06,Bambi20}. 

Many studies suggest that the cool accretion disc extends down to a few Schwarzschild radii ($R_{\rm S}=2GM_{\rm BH}/c^2$, where $G$ is gravitational constant, $M_{\rm BH}$ is the BH mass and $c$ is the speed of light), i.e. at the innermost stable circular orbit \citep[ISCO; e.g.,][]{Toms08,Reis10,Garcia15} in the hard state, while a large number of other works find that in this state the accretion disc is truncated at a larger radius, tens of Schwarzschild radii \citep[e.g.,][]{GD04_theory,DeMarco15reverber,DeMarco16reverber,Mahmoud19}. 
The main techniques used to infer the inner radius of the disc include spectral fitting of the line shape \citep[sensitive to the truncation radius; for a review see][]{Bambi20}, assessment of the temperature-luminosity relation of the soft component \citep[assumed to be coming from either a standard or an irradiated accretion disc, see review in][]{DGK07}, and timing properties of the iron line and the thermal soft emission \citep[the so-called thermal reverberation; see][for a review of this method]{uttley14reverberation}.

All these methods have their shortcomings and in many cases the inner radii inferred by different methods do not match even for the same data set \citep[e.g.,][]{KDD14,Garcia15}. 
Emission at soft energies is affected by the Galactic absorption, making the determination of the exact spectral shape difficult and potentially leading to inaccurate estimates of luminosity and temperature of the disc.
Interpretation of timing results depends on the pre-assumed contribution of different spectral components, so that the lags of soft energies with respect to the harder ones are typically ascribed to thermal reverberation.
In addition, studies are also often limited by either low photon statistics or instrumental imperfections \citep[such as pile-up effects, see discussion in][]{diaz-trigo_done10}.

\begin{figure*}
   \centering 
   \includegraphics[width=0.95\textwidth]{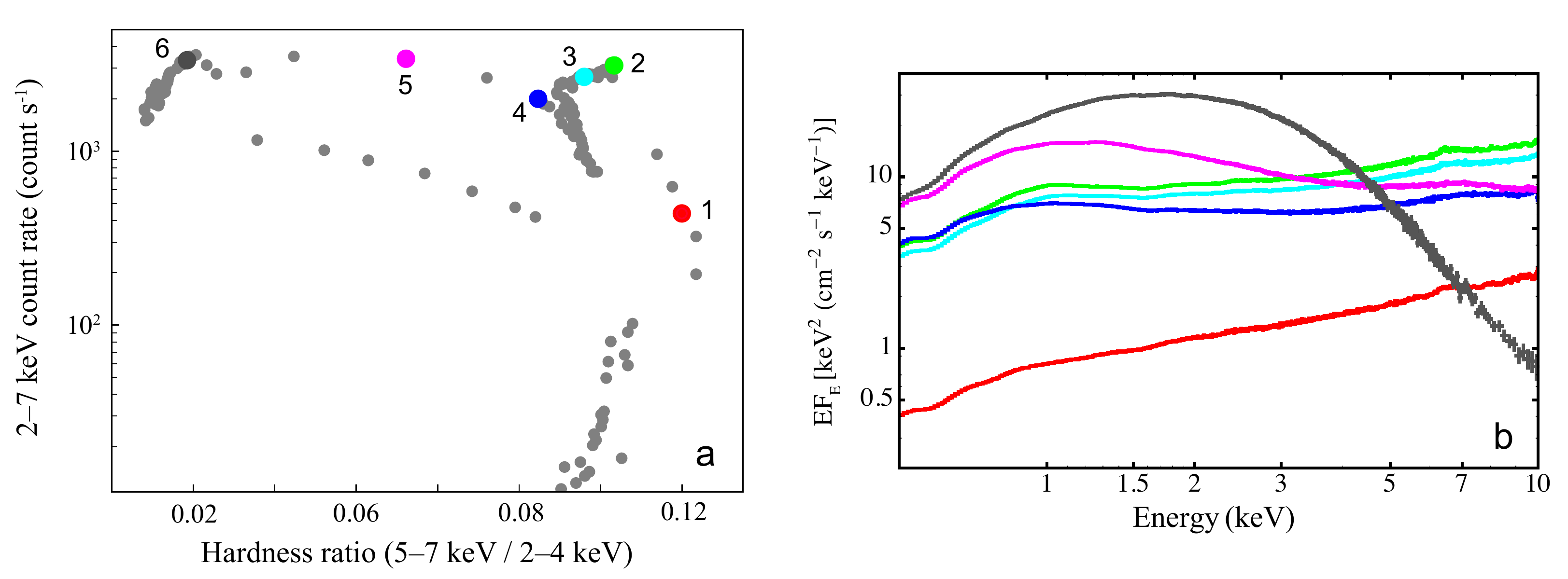} 
     \caption{Left panel: hardness-intensity diagram showing the location of the observations analysed in this paper. Numbers correspond to the Epochs in Table~\ref{tab:obs}.
    Right panel: comparison of spectra for all Epochs: 1 (red), 2 (green), 3 (cyan), 4 (blue), 5 (magenta) and 6 (gray). All are deconvolved against a power law with photon index $\Gamma=2$.}
\label{fig:HRplotspectra}
\end{figure*}

Fourier frequency-resolved spectroscopy serves as an independent diagnostic to probe the geometry of the black hole vicinity.
The technique has previously been applied to accreting neutron stars, where it helped discriminating between the disc and the boundary layer spectra \citep{GRM03,RG06}.
Applications to black holes included extraction of broadband spectra at various ranges of aperiodic variability frequencies \citep{RGC99,RGC01,GZ05,uttley11,cassatella12,AHD13,AD18}, or at harmonics of the quasi-periodic oscillations \citep{ADH14,AD16}.
An important assumption is the high coherence of variability in different energy channels, as, for instance, interplay between two components varying out of phase may lead to interference patterns in power spectra and mimic the reduction/enhancement of variability at certain frequencies \citep{V16,V18}.

High coherence is expected to be achieved in energy ranges with spectral features, such as the iron line complex.
Early studies \citep{RGC99,gilfanov00,RGC01} showed that the dependence of the iron line equivalent width (EW) on the Fourier frequency can be used to infer the disc transfer function.
These studies found a clear drop of EW at higher frequencies, which could be explained either by the finite size of the reflector or that the emission from the innermost regions, where rapid variability is assumed to originate, does not strike the disc. Both cases require the accretion disc to be truncated at some distance from the black hole. 
Albeit being robust, the technique can only be applied to bright targets, such as Cygnus~X-1 and GX~339--4, and using an instrument free of pile-up effects.
Both conditions have been satisfied for the exceptionally bright BH XRB MAXI~J1820+070 as viewed by the Neutron star Inner Composition Explorer ({\nicer}) instrument \citep{gendreau16} mounted on the International Space Station.

The black hole X-ray binary MAXI J1820+070 was detected in X-rays starting in March 2018 \citep{Kawamuro18}. It rapidly brightened, reaching a peak flux of about 4 Crab \citep{Shidatsu19ATel}, and proceeded to undergo a complete outburst cycle before again nearly reaching quiescence in February 2019 \citep{Russell19ATel}. The high brightness of the source combined with the low column density \citep[$N_{\rm H}\sim10^{21}$\,cm$^{-2}$;][]{Uttley18ATel}, have led to monitoring campaigns by many instruments, including \nicer\,, allowing detailed studies of the evolution during the outburst.

In this paper we analyse observations taken by \nicer\, during the rising phase of the outburst. 
In contrast to previous studies, we track the changes of spectra and timing properties throughout the whole transition to the soft state.
We study the evolution of the temporal properties as a function of energy and frequency, iron line shape and dependence of its EW on frequency -- all being tracers of the inner accretion geometry. 
We discuss the results in the context of two alternative geometries: truncated disc-inner flow and lamp-post corona with non-truncated disc.

\begin{table}
    \centering
    \begin{tabular}{l l l l}
    \hline
    \hline
        Epoch & {\it NICER} ObsID & Obs date & Exposure (ks) \\
        \hline
        1 & 1200120103 & 2018-03-14 & 9.5 \\
        2 & 1200120106 & 2018-03-21 & 5.4 \\
        3 & 1200120130 & 2018-04-16 & 10.6 \\
        4 & 1200120194 & 2018-07-03 & 5.0 \\
        5 & 1200120196 & 2018-07-05 & 3.6 \\
        6 & 1200120210 & 2018-07-18 & 1.9 \\
    \hline
    \end{tabular}
    \caption{Summary of {\it NICER} observations analysed in this work.}
    \label{tab:obs}
\end{table}

\section{Data}

The {\nicer} data have been reduced using the \texttt{NICERDAS} tools included in the \texttt{HEASoft} software package version 6.28. We applied standard screening criteria and the instrument background for each observation was estimated using \texttt{nibackgen3C50} tool (Remillard et al., in prep.) supplied by the {\nicer} Guest Observer Facility.\footnote{\url{https://hera.gsfc.nasa.gov/docs/nicer/analysis_threads/background/}} Response files were retrieved from the HEASARC Calibration Database (CALDB) system, version 20200727. The \textit{NICER} response has undergone a series of improvements since launch, leading to noticeable changes when comparing spectra for different responses. 

In order to determine the position of each observation in the hardness-intensity diagram, a hardness ratio was determined as the ratio of count rates in the 5--7 and 2--4~keV bands. Observations were then chosen to span the rising phase of the outburst. Table~\ref{tab:obs} shows a summary of the observations.  

To study the variability, power spectra were calculated from the light curves using the HEASoft tool \texttt{powspec}. The normalization was chosen so that the integral of the power spectrum gives the fractional root-mean-square variability \citep[fractional rms;][]{miyamoto91}. At high count rates, telemetry saturation may cause fragmentation of the data. In order to avoid contamination of the power spectra, we have screened the good time intervals (GTIs) for each event file so that only complete light curve segments are used in calculating the power spectrum.

\section{Analysis and results}

A hardness-intensity diagram of all \nicer\, observations of MAXI J1820+070 is shown in Fig.~\ref{fig:HRplotspectra}a. 
The datasets considered in this work are indicated by coloured circles.
We note that the analysed observations were chosen to span a large range of hardness and cover the complete evolution of the source from hard to soft state. The first observation (Epoch~1) is one of the first {\nicer} observations during the outburst, when the source was in the low/hard state. At the other end is the last observation (Epoch~6), when the source has completed the transition to the soft state. 

\begin{figure*}
   \centering
  \includegraphics[height=0.26\textwidth]{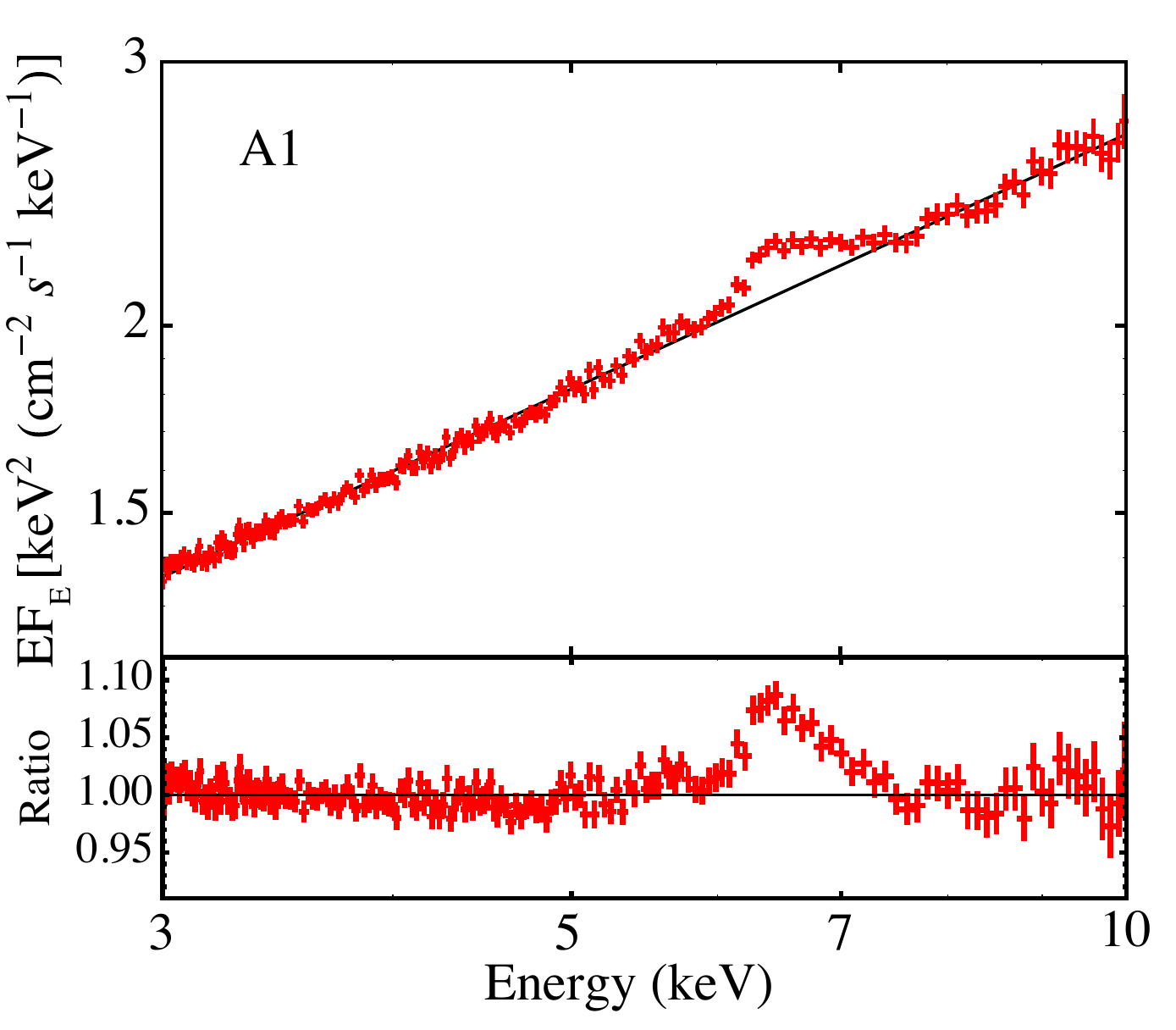}
   \includegraphics[height=0.26\textwidth]{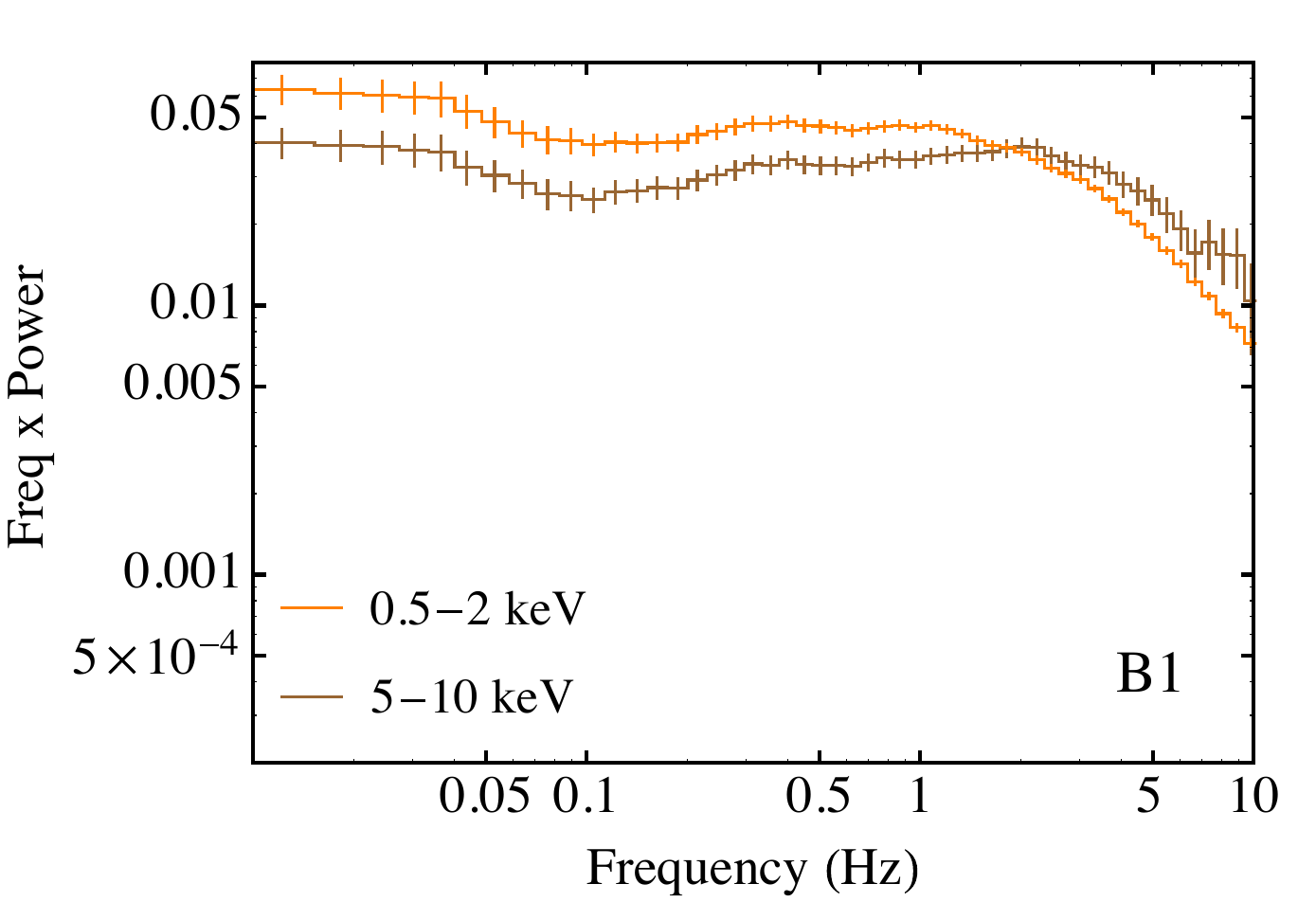}
    \includegraphics[height=0.26\textwidth]{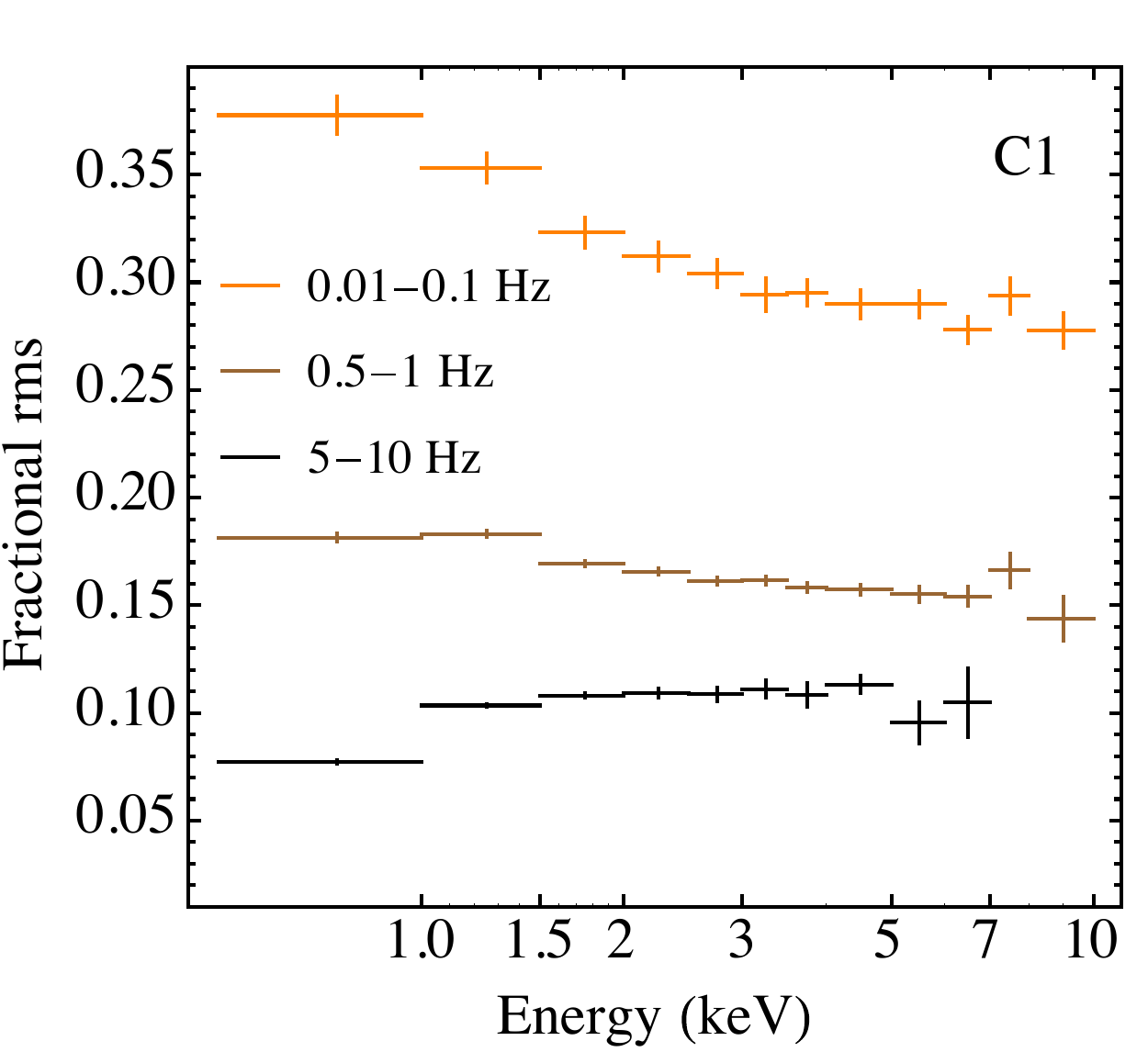}
    \\
 \includegraphics[height=0.257\textwidth]{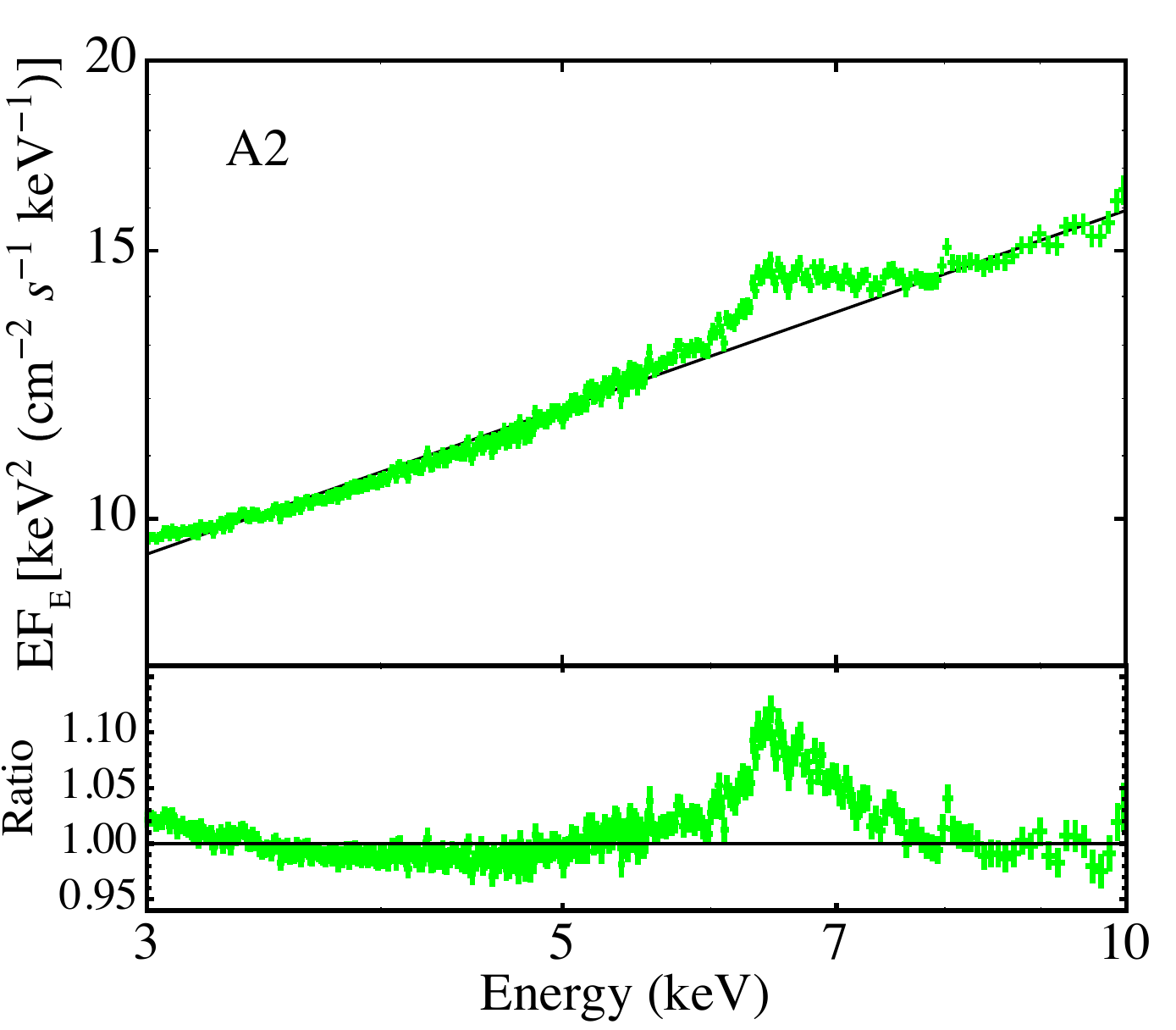}
   \includegraphics[height=0.26\textwidth]{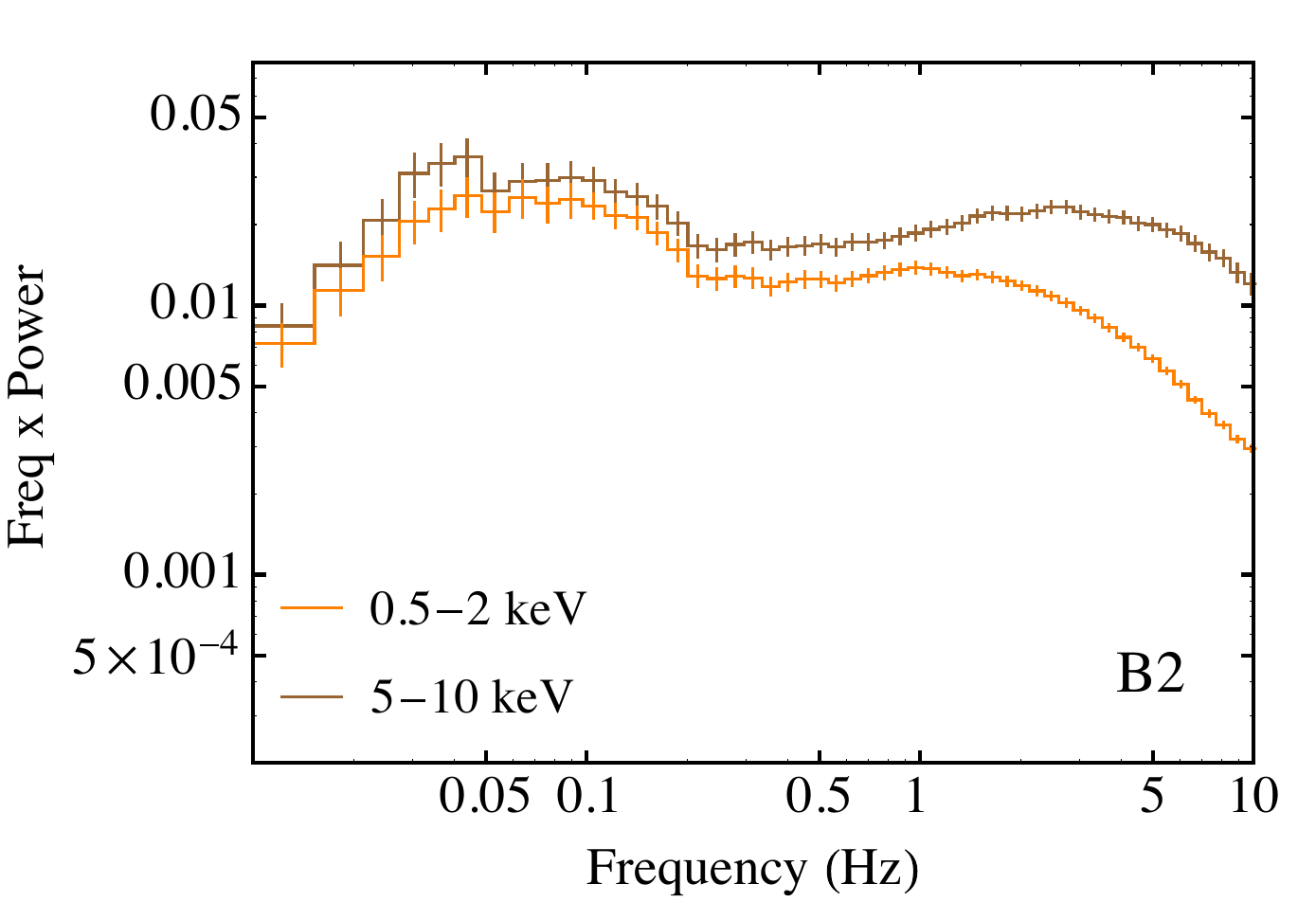}
    \includegraphics[height=0.26\textwidth]{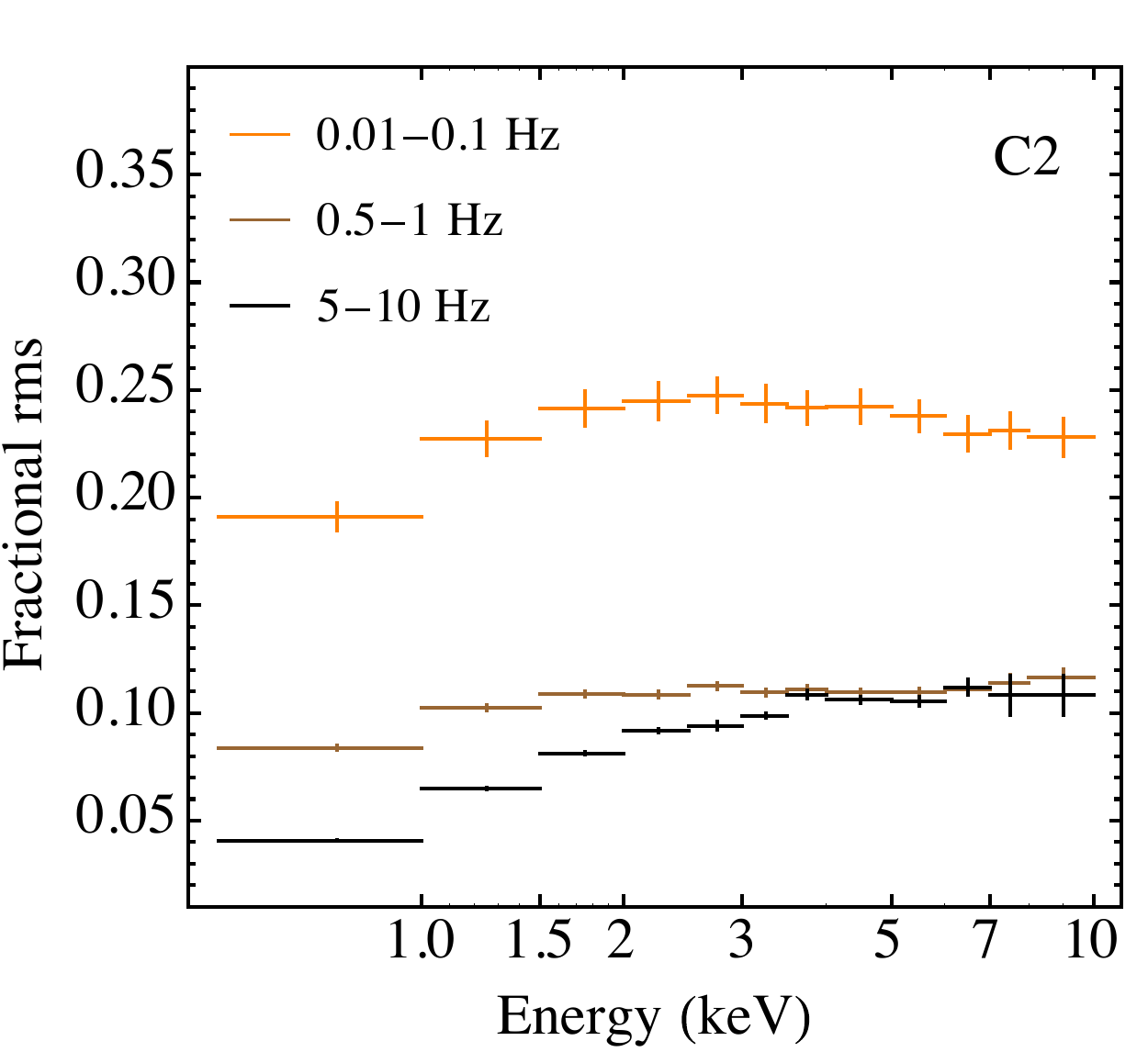}
    \\
 \includegraphics[height=0.257\textwidth]{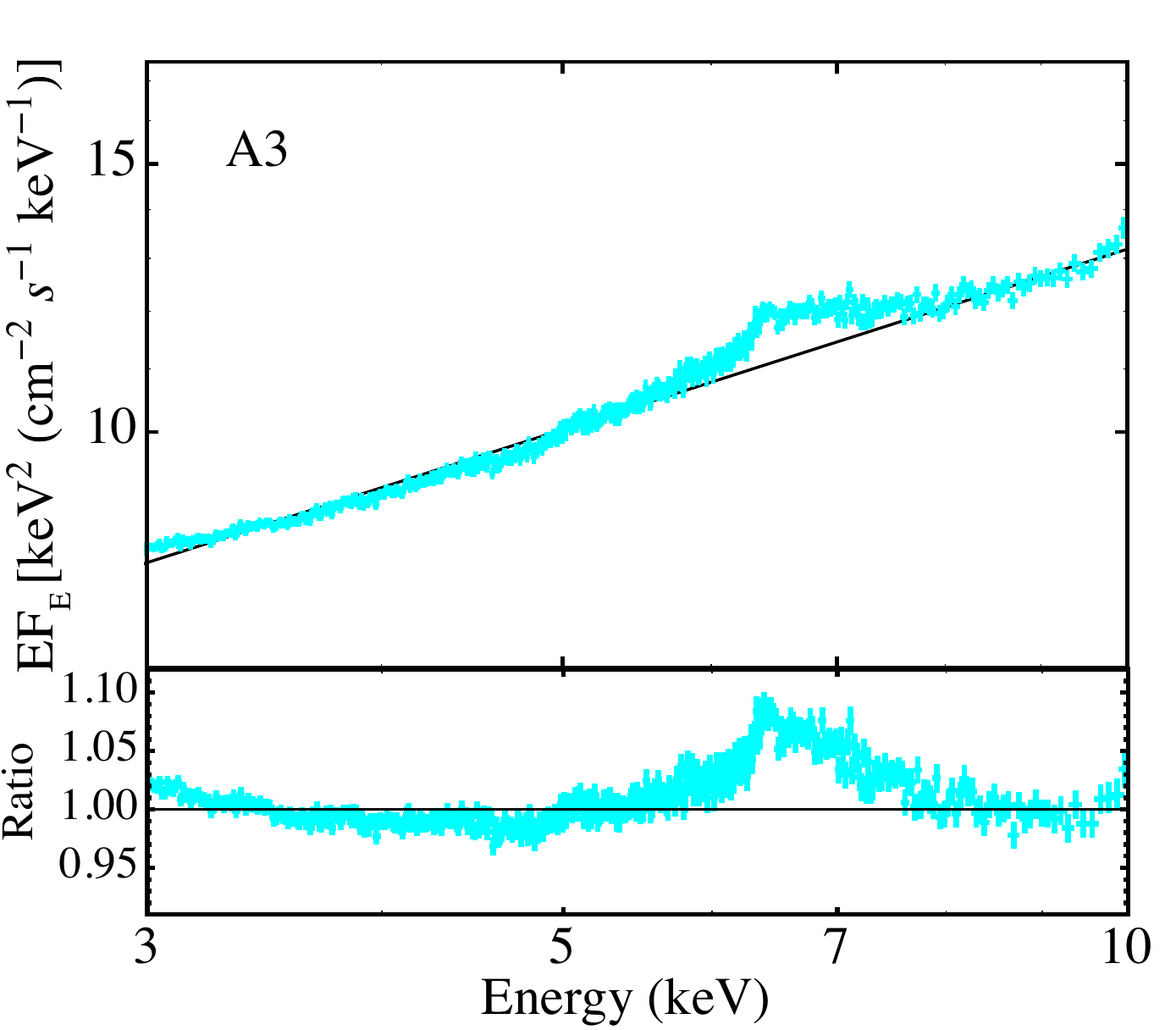}
   \includegraphics[height=0.26\textwidth]{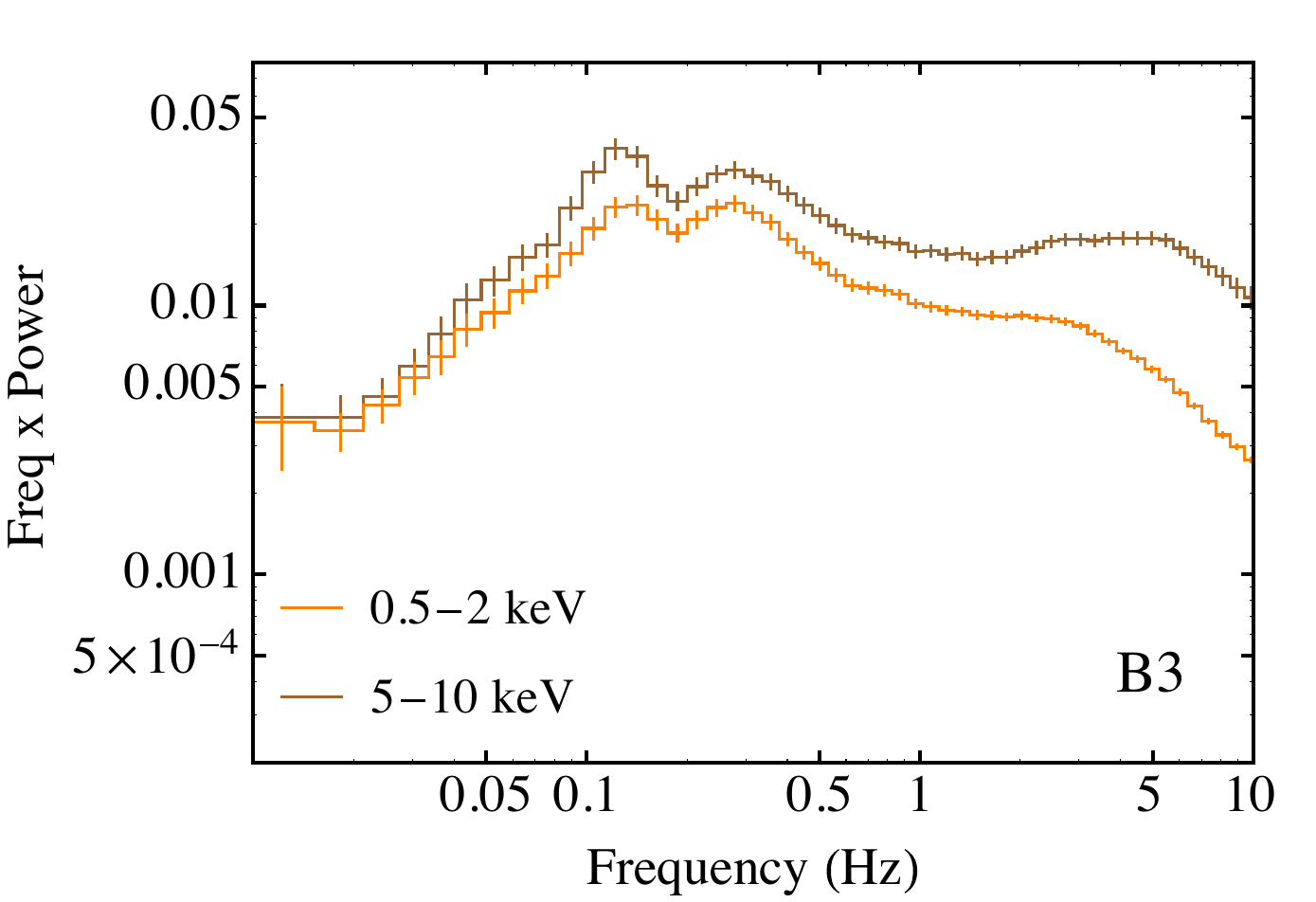}
    \includegraphics[height=0.26\textwidth]{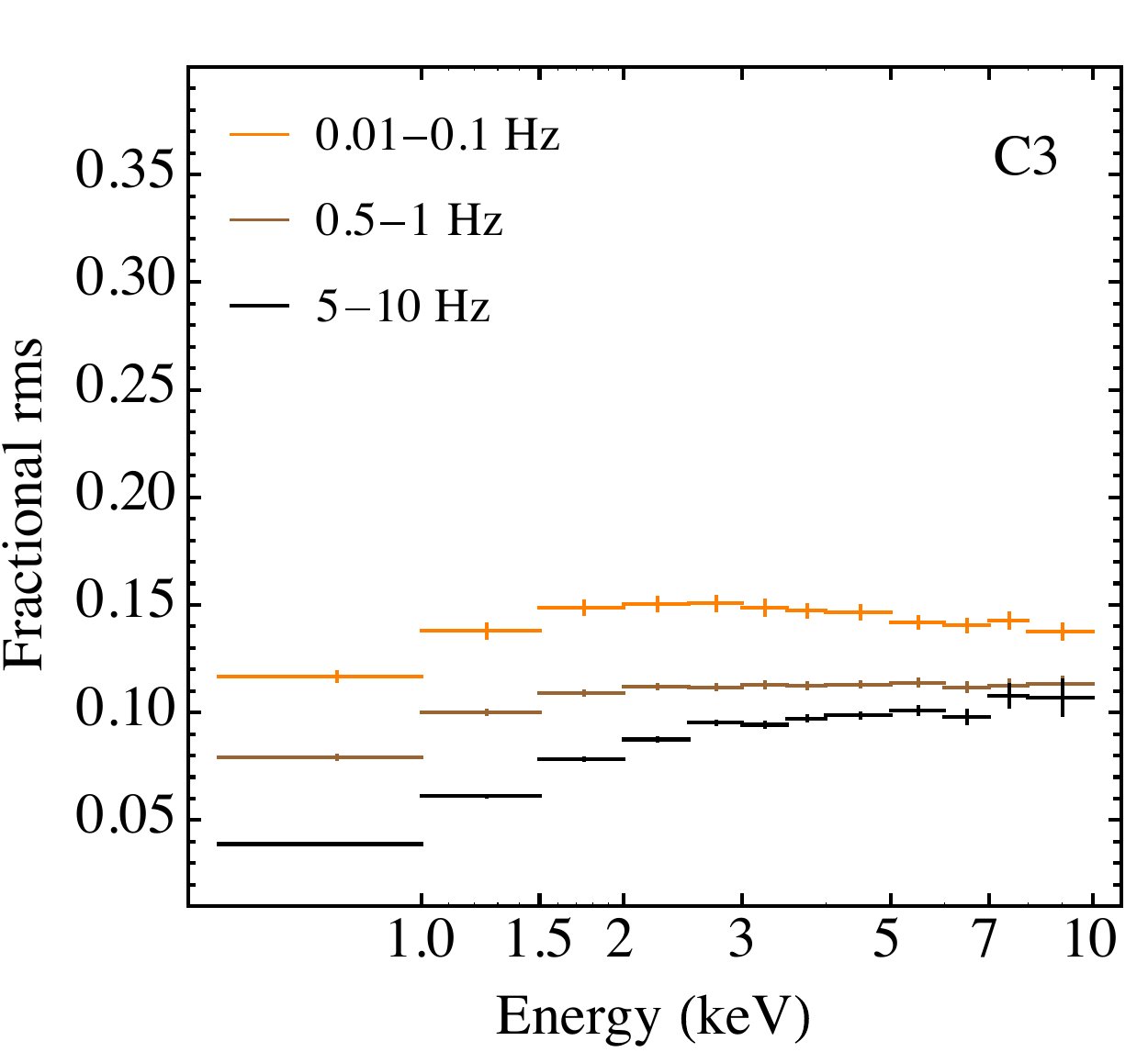}
    \\
 \includegraphics[height=0.26\textwidth]{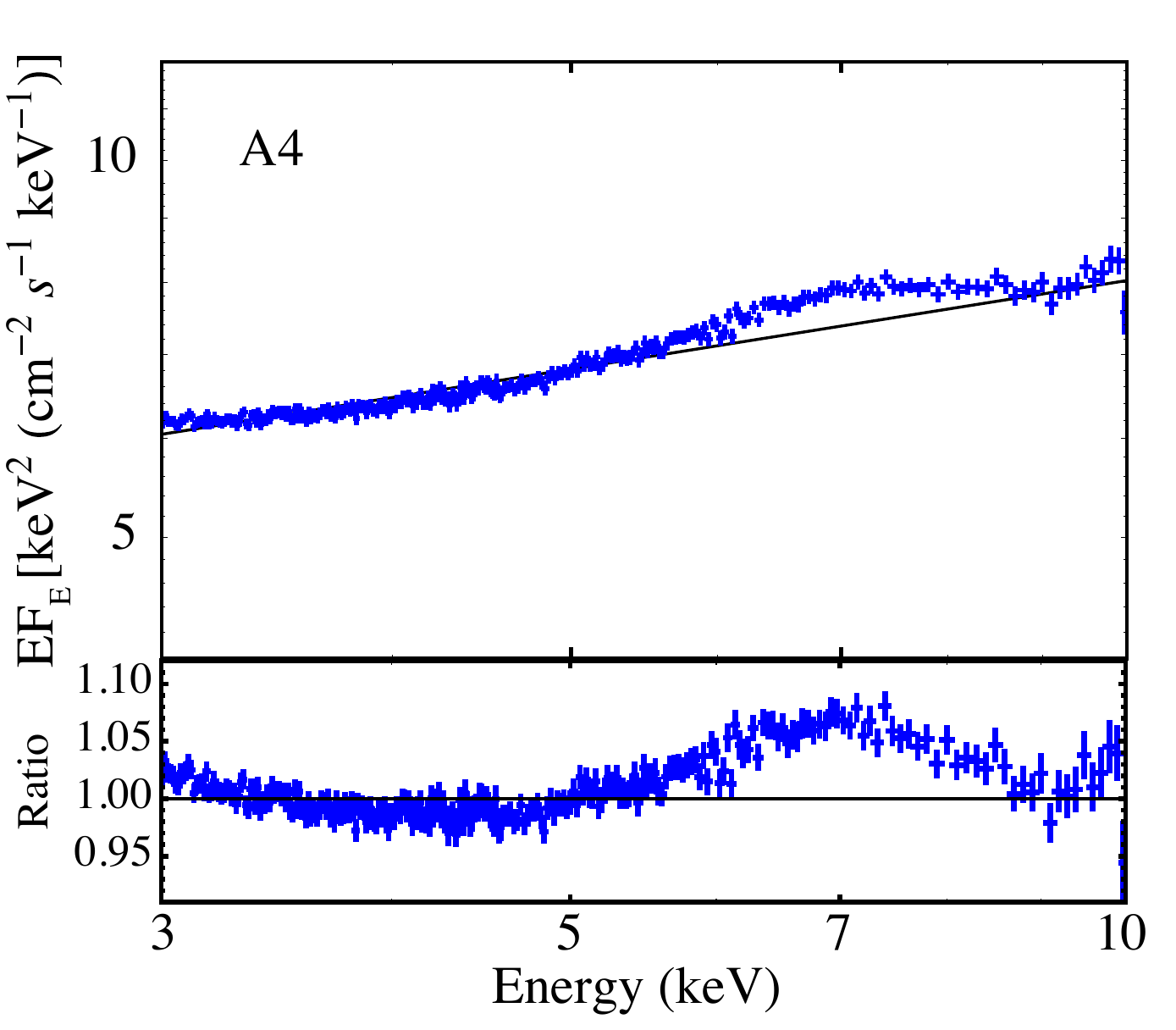}
   \includegraphics[height=0.26\textwidth]{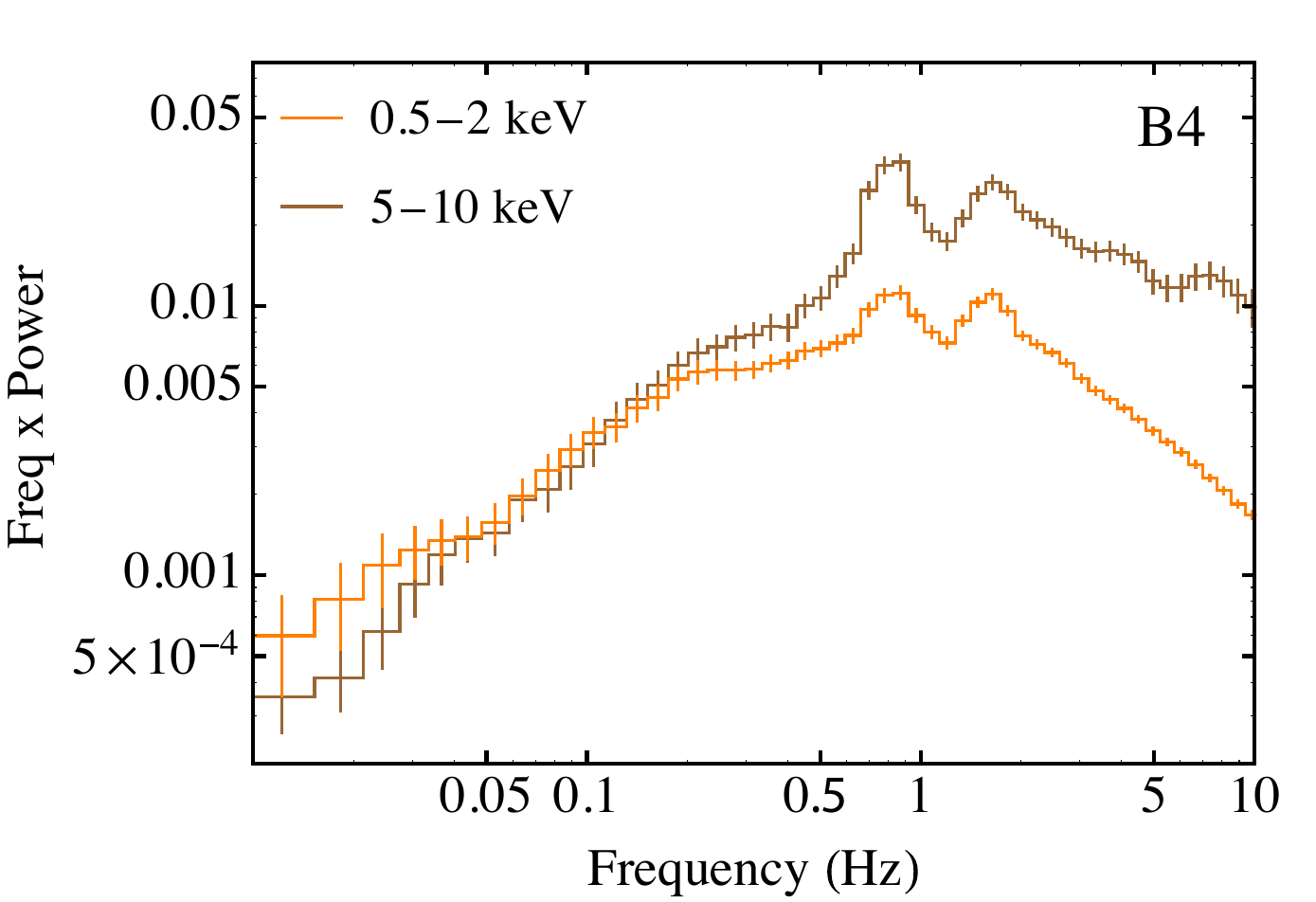}
    \includegraphics[height=0.26\textwidth]{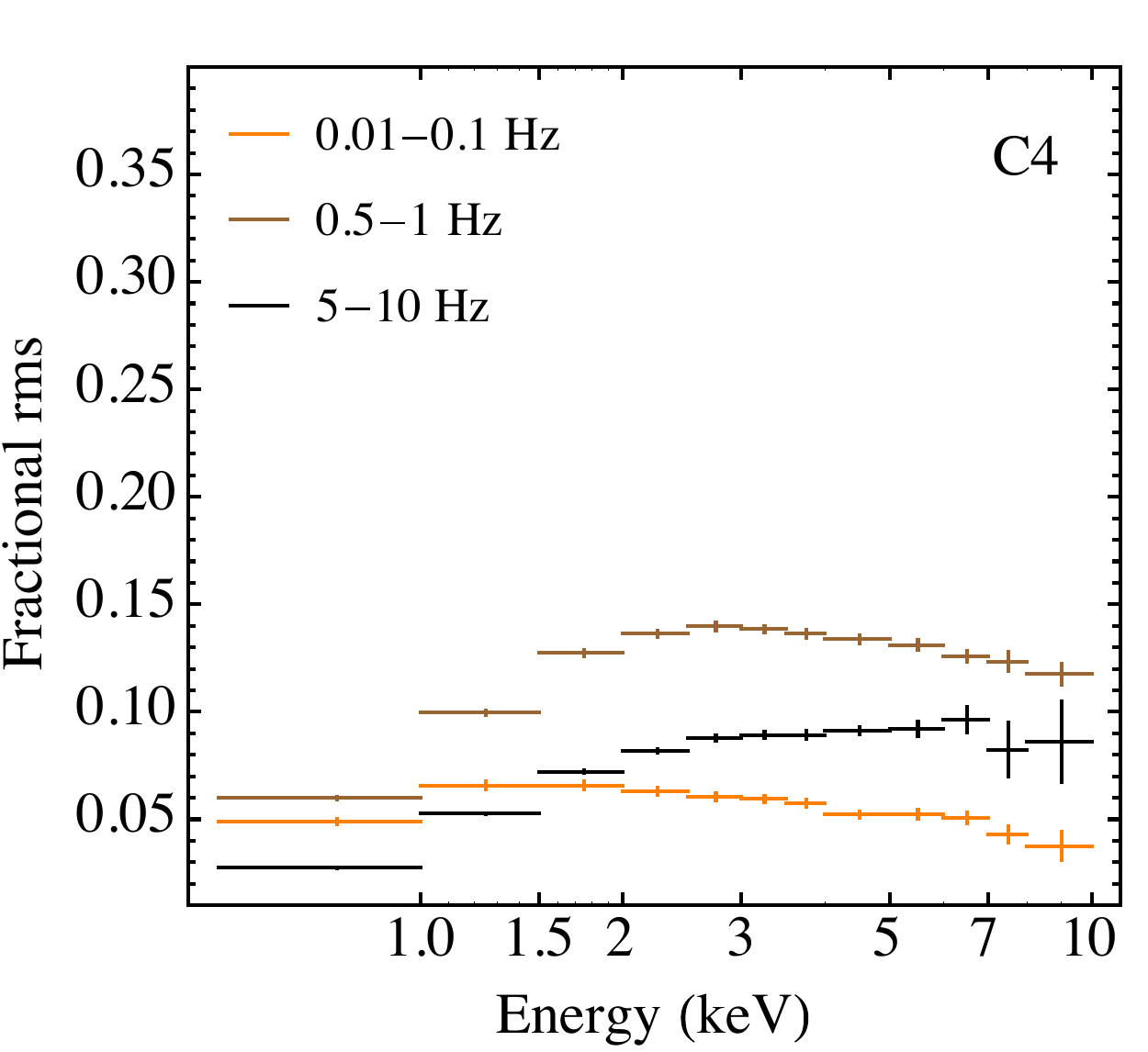}
    \\  
\includegraphics[height=0.257\textwidth]{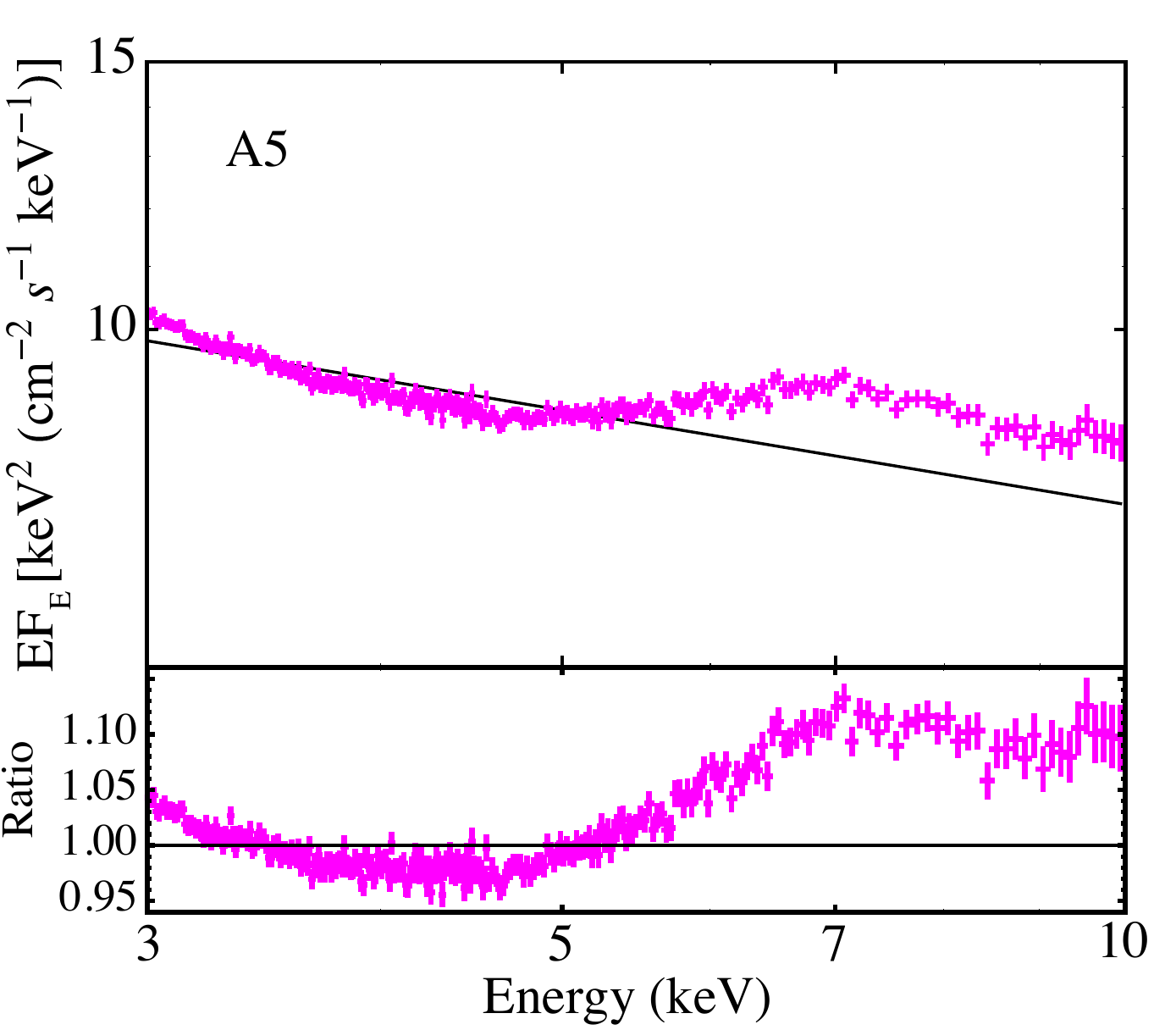}
   \includegraphics[height=0.26\textwidth]{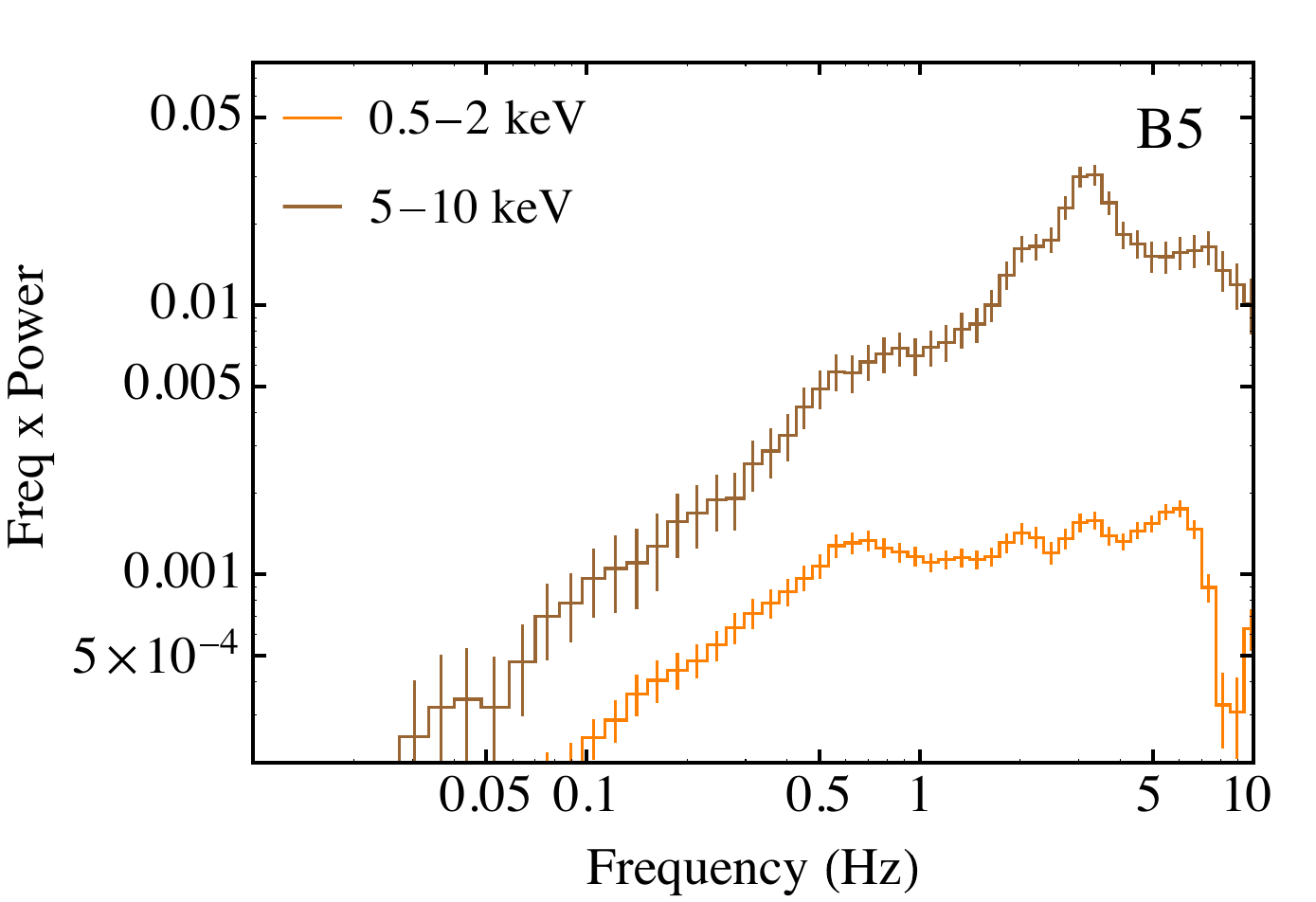}
    \includegraphics[height=0.26\textwidth]{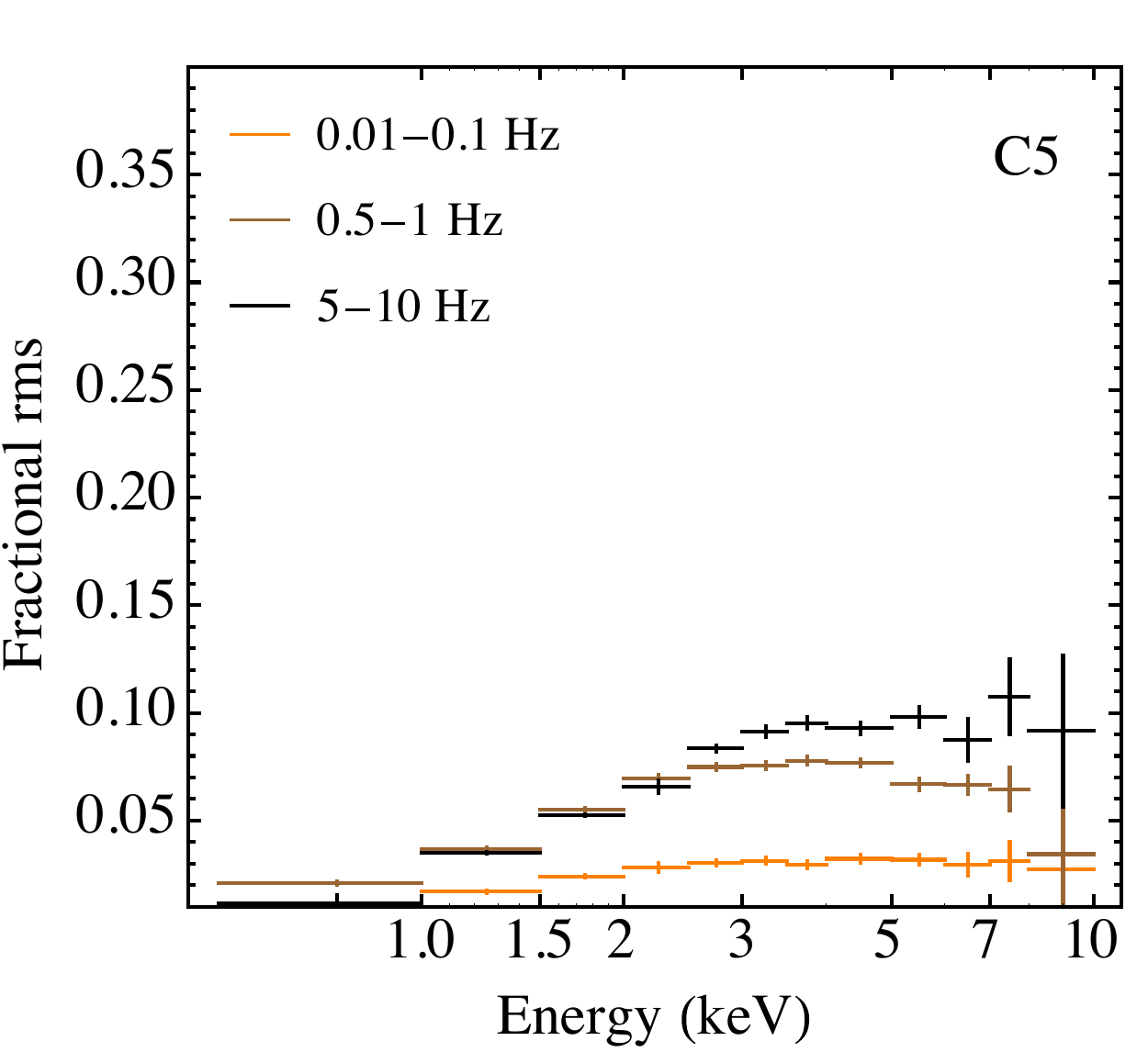}
    \\
  \caption{Comparison of 3$-$9~keV spectra (column A), power spectra (0.5$-$2~keV and 5$-$10~keV; column B) and rms vs energy in three different frequency bands (column C), for Epoch 1 to 5. 
} 
\label{fig:comparison}
\end{figure*}

Fig.~\ref{fig:HRplotspectra}b shows the spectrum of the observations all deconvolved against a power law with photon index $\Gamma=2$. The figure clearly shows the change in overall flux as well as the spectral shape.  
As seen in Fig.~\ref{fig:HRplotspectra}b, in the early stages of the outburst the spectrum is dominated by a power-law component over the entire energy range, generally interpreted as a Comptonization continuum \citep{ZG04}. As the outburst evolves, with flux increasing and hardness decreasing, a second component starts to appear at lower energies. This component is likely connected to thermal emission from the accretion disc \citep{SS73}. As the source moves into a softer state, both the strength and temperature of this component increase, and in the softest observation (Epoch~6) it dominates the spectrum.
In order to probe the connection between the spectral evolution and the accretion geometry, we also consider the temporal evolution and employ the frequency-resolved spectroscopy technique. We start by summarizing the spectra, power spectra and rms-energy evolution for Epochs 1 to 5 in Fig.~\ref{fig:comparison} (Epoch~6 is not included in detailed timing analysis because of the low level of variability, as expected for the disc-dominated state).

\subsection{Power spectra and rms-energy dependence}

We compute the power spectra in two energy bands, 0.5--2 and 5--10\,keV.
These are shown in column (B) of Fig.~\ref{fig:comparison}.
These power spectra clearly show an evolution of the temporal behaviour. As the source softens, there is an overall shift of power towards higher frequencies and an appearance of the quasi-periodic oscillations (QPOs), both of which are common features in X-ray transients \citep{RM06}. Comparing the two energy bands allows us to find also other trends. 

In Epochs~2 to 4, the power spectra in the two energy bands are quite similar, but the lower-energy band is suppressed at higher frequencies. The relative suppression of variability in the 0.5$-$2\,keV band increases as the spectrum becomes softer.
In the final observation (Epoch~5; B5) nearly all variability lies in the 5$-$10\,keV band, and the lower-energy power is suppressed.
An interesting behaviour is detected in Epoch~1 (panel B1), where the variability at lower energies dominated over the variability at higher energies for frequencies up to 2~Hz (an indication of power excess for lower energies might also be seen in panel B4, below 0.05~Hz).
Similar behaviour, with lower energies having higher power at lower frequencies, has been reported for Cyg~X-1 \citep{RIvdKcygx1}.

To further study the variability, we compute the rms in three frequency ranges: 0.01$-$0.2, 0.5$-$1 and 5$-$10\,Hz.
The right column (C) in Fig.~\ref{fig:comparison} shows the rms in these ranges as a function of energy. 
Throughout the datasets, the highest frequency range rms (black points) is an increasing function of energy, with fractional rms of $\sim10\%$ above 5\,keV.  
On the other hand, the lowest frequency range rms (orange points) declines from a fractional rms of $>30\%$ in Epoch~1 (C1) to $<5\%$ in Epoch~5 (C5). Column C also allows us to see how the suppression of variability at low energies first appears in the lowest frequencies (orange points; panels C1 to C5), then later in the outburst also affecting the middle frequency range (brown points; panels C4 and C5). Panels B5 and C5 together show that in this state variability is suppressed at frequencies below 0.05\,Hz or energies below 1\,keV.

\begin{figure}
  \includegraphics[width=0.45\textwidth]{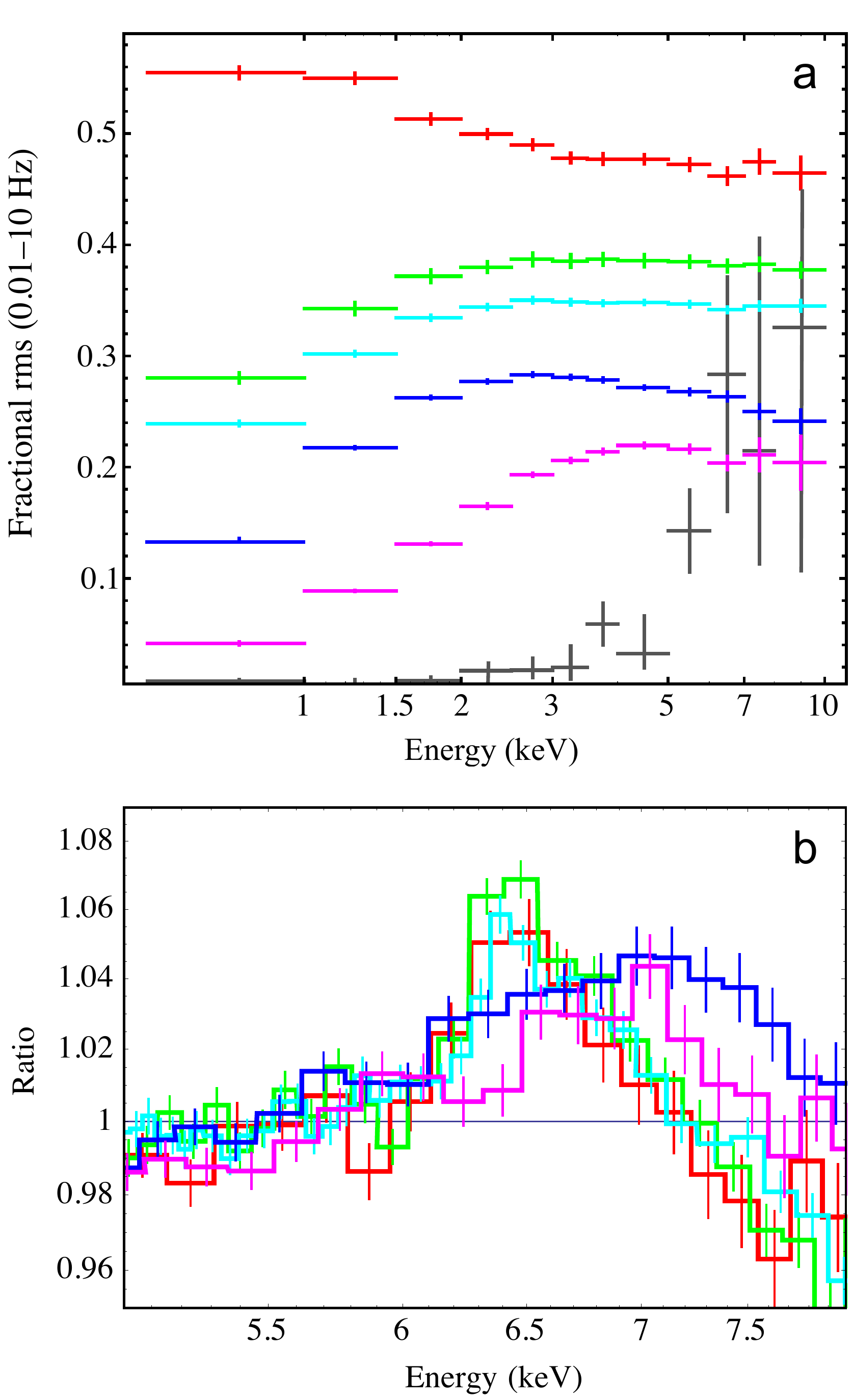}
     \caption{Upper panel: Evolution of rms vs energy for all Epochs. From top to bottom: Epoch~1 (red), 2 (green), 3 (cyan), 4 (blue), 5 (magenta) and 6 (gray). 
     Lower panel: Evolution of the iron line shape throughout the state transition.
     Residuals compared to a power-law fit in the 4--10~keV band for Epochs 1 (red), 2 (green), 3 (cyan), 4 (blue) and 5 (magenta).
}
 \label{fig:rmslinecomp}
\end{figure}

The evolution of rms-energy dependence during the outburst can be seen in Fig.~\ref{fig:rmslinecomp}a, where the rms in the 0.01$-$10\,Hz range as a function of energy is shown for all the analysed observations (also including Epoch 6 where the overall rms is very low). 
The rms decreases as the source evolves to softer states, which is a general trend for XRBs and has been reported in several previous studies \citep[e.g.,][]{churazov01,grinberg14}. Comparing with Fig.~\ref{fig:comparison} it becomes clear that the changes are primarily tied to the lower frequencies. We also find a change in energy dependence: initially the rms increases towards lower energies, but at later stages in the outburst it is instead suppressed  \citep[similar to previous findings, e.g.][]{GZ05,uttley11,cassatella12,DeMarco15}.
Although it does not reach statistical significance, there is a weak indication of the rms being systematically lower around the energy of the iron line.

The rms in the soft state observation (Epoch~6) is below 5\% up to $\sim5$\,keV, as commonly reported for the disc-dominated state in XRBs \citep{vdK06}. At higher energies there is a tendency for the variability to increase; however, we caution that the count rate is low and the uncertainties are large. 

\begin{figure}
   \centering 
   \includegraphics[width=0.45\textwidth]{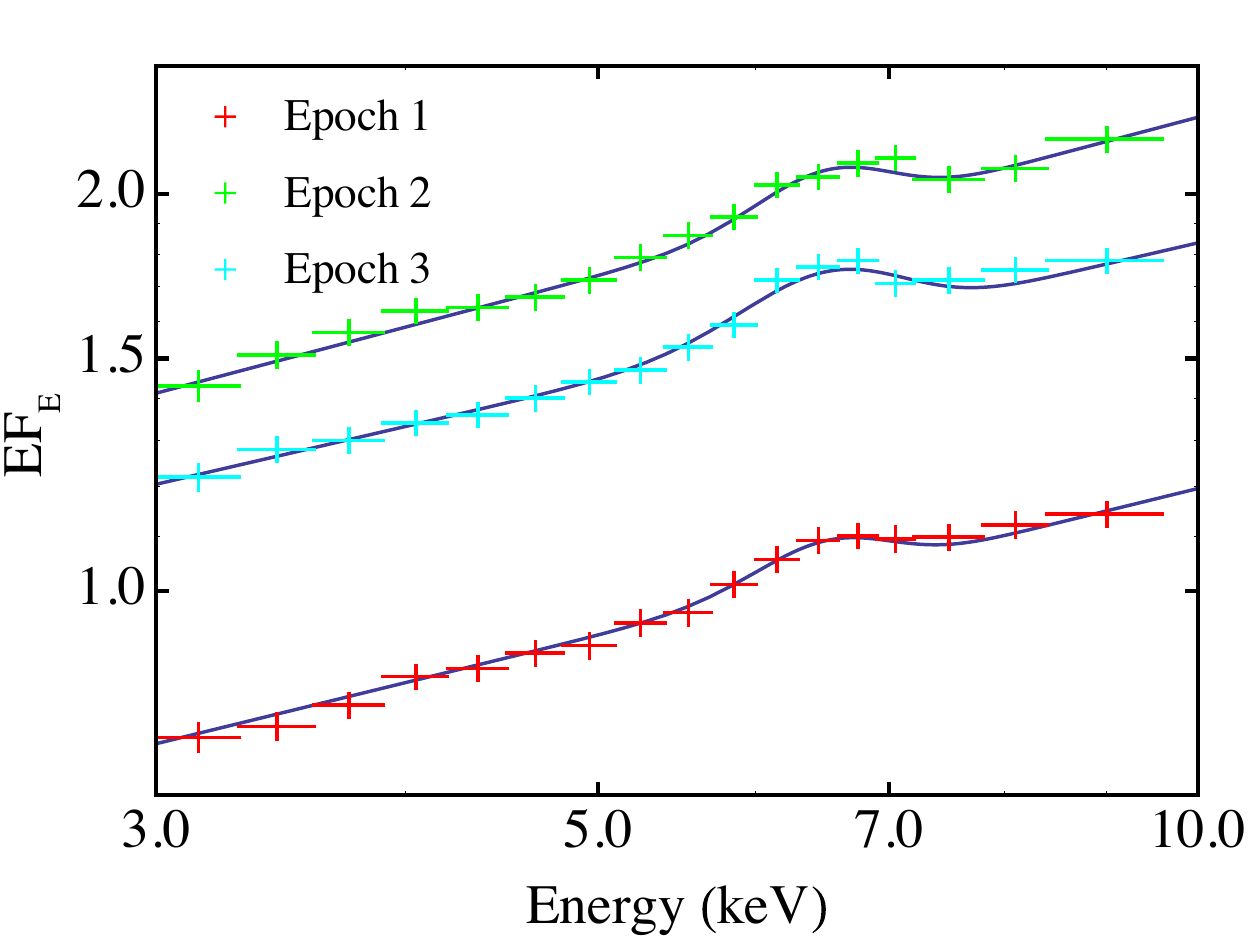}
     \caption{Examples of the frequency-resolved spectra from which the EW of the line is derived. The solid lines indicate the model comprising a power law and gaussian line. The frequency ranges for the shown spectra are 0.01-0.02 Hz for Epoch 1 (shifted upward by a factor of 3), and 0.4-0.8 Hz for Epochs 2 and 3.} 
\label{fig:ewfreq_fits}
\end{figure}

\subsection{Iron line shape}
 
An apparent feature in all spectra in Fig.~\ref{fig:HRplotspectra}b is the iron line around 6.4~keV, arising as hard emission is reprocessed in the accretion disc \citep{basko74,george91}. It is most prominent in the spectra of the initial stages of the outburst, but appears to be present also at later stages.
As the source transits to softer states, the shape of the line is expected to change in response to many factors: changing ionization of the cool disc material, inner disc radius as well as the shape (power-law slope) of the incident X-ray continuum may all affect the width of the line and its peak energy.

Excess emission (to the Comptonization continuum) between 5 and 7~keV originates from the K$\alpha$ lines of iron at different levels of ionization, as well as from the photoionization edges. 
We assume that the line emission completely outshines the possible contribution from the edges.
We fit a power law to each spectrum, which represents the underlying continuum, ignoring the 5.5--7.5\,keV range.
The ratio of the data to this model gives the excess emission, which can be attributed to the fluorescent line. 
While it is clear that the spectrum has some curvature even in this narrow range, the approach allows us to at least qualitatively study the iron line. The results are shown in the left column (A) of Fig.~\ref{fig:comparison}.

The figures clearly show changes in strength and shape of the iron line region as the spectral state of the source changes. In the first observation (Epoch~1; panel A1), the line is not so strong and the peak is less than 10\% above the power-law model. In the following observations, the line grows stronger and also appears to broaden towards higher energies. In the last observation (Epoch~5; panel A5) the line complex appears to extend above 9\,keV. However, we believe this is more likely to be due to the power law being a poor representation of the underlying continuum -- as can be seen from the overall spectrum there is significant curvature likely caused by the disc component becoming stronger and influencing the spectrum above 3\,keV. 

To better see the evolution of the line, we perform a power-law fit to the 4--10\,keV range without exclusions, and show the residuals for Epochs 1 to 5 in Fig.~\ref{fig:rmslinecomp}b. In the first three epochs, the main changes to the line are in the core and blue wing. As the source reaches softer states (Epochs 4 and 5), there is a clear change to the centroid of the line, which is found to shift to higher energies. As the high-energy emission ($>7$~keV) substantially decreases between Epoch~2 and 5 (the observed 7--10\,keV flux is $\sim40\%$ lower in Epoch 5 compared to Epoch 1), the increase in line centroid is unlikely to be caused by an increase in disc ionization.

\begin{figure*}
   \centering 
   \includegraphics[width=0.3\textwidth]{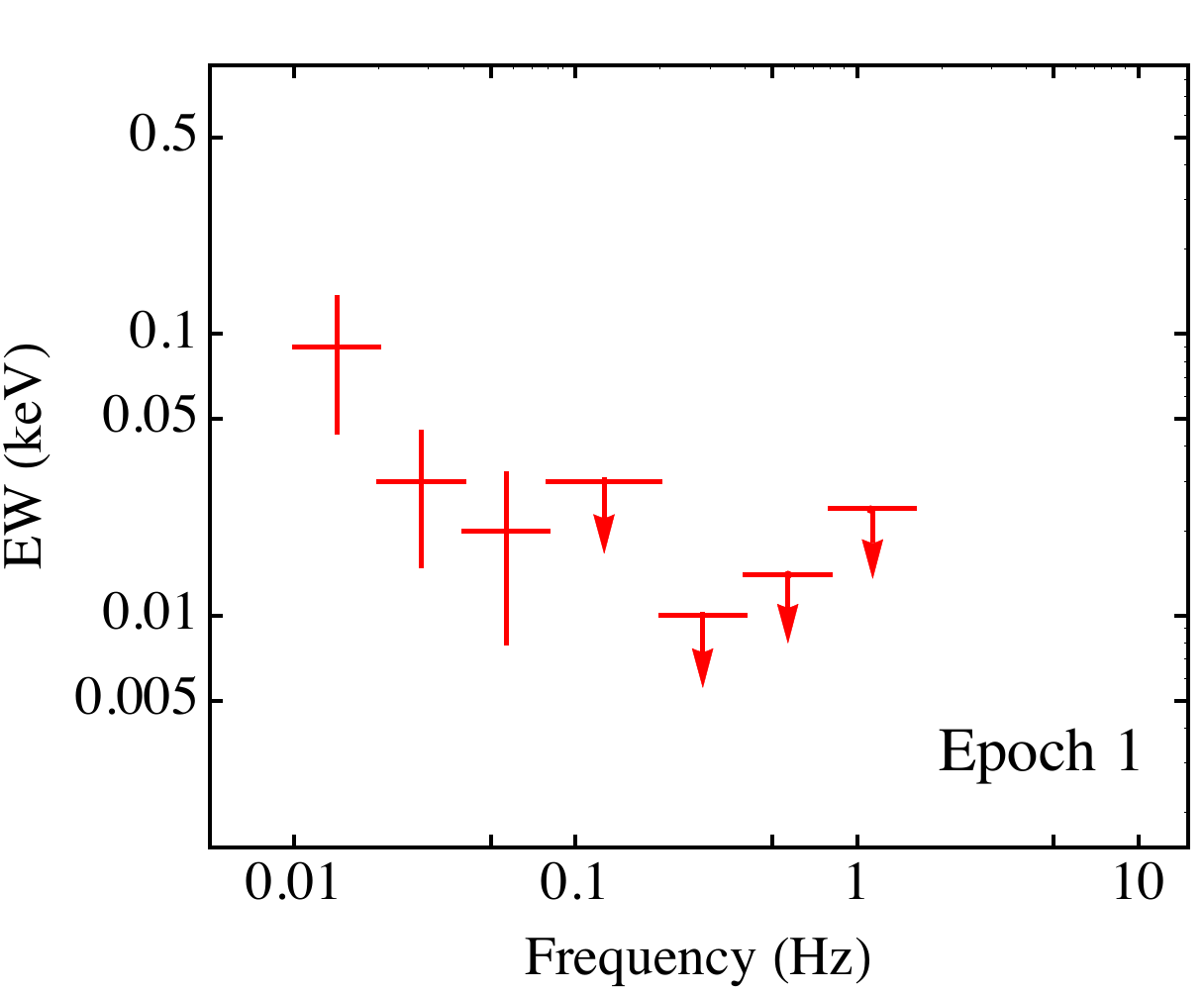}
   \hspace{0.5cm}
  \includegraphics[width=0.3\textwidth]{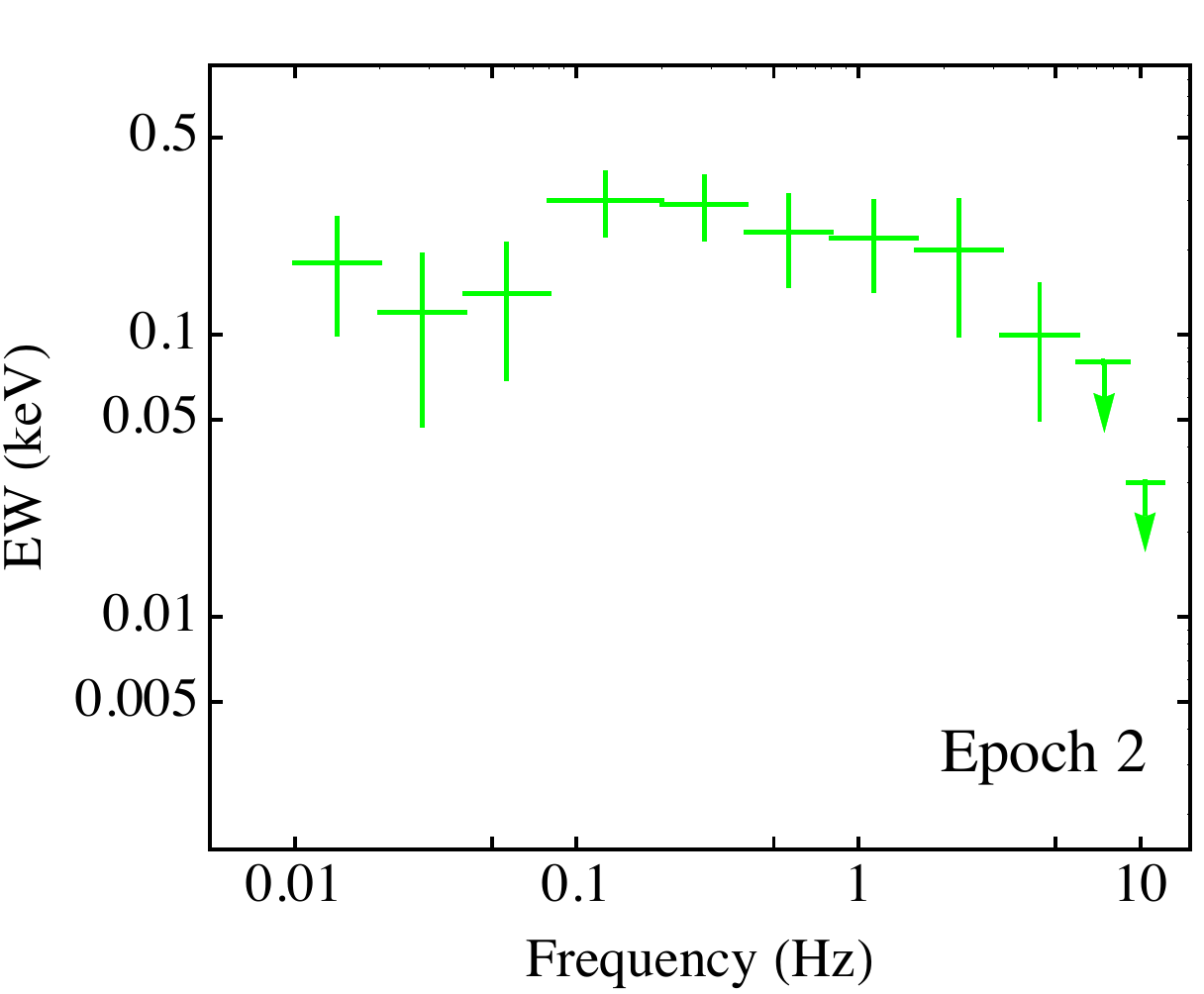}
  \hspace{0.5cm}
  \includegraphics[width=0.3\textwidth]{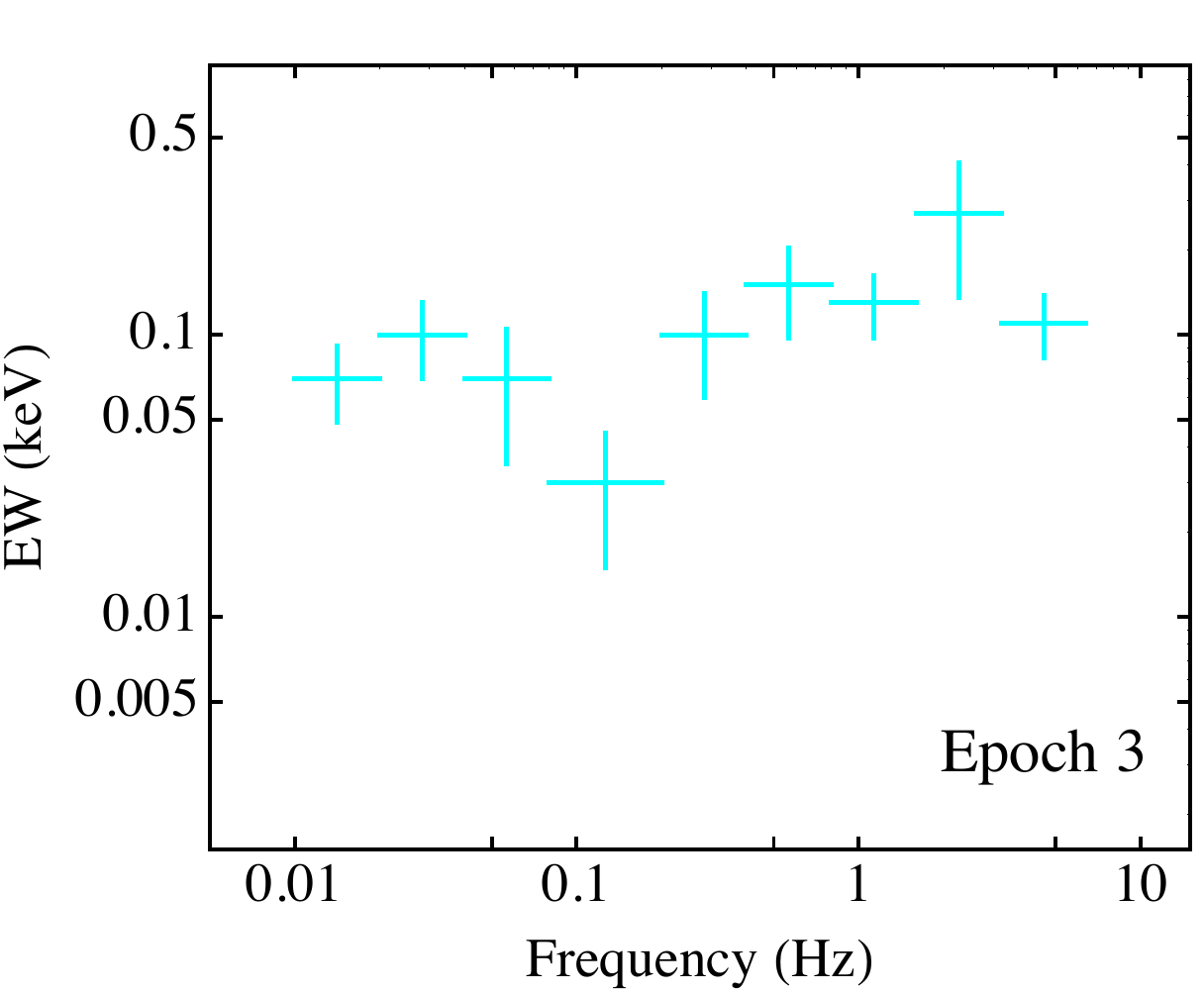}
     \caption{EW dependence on Fourier frequency for Epochs 1, 2 and 3. In the first two epochs there is a clear cut off at higher frequencies, while no such feature is seen in Epoch 3.} 
\label{fig:ewfreq_newresp}
\end{figure*}

\subsection{Variability of the iron line}

As the iron line arises through reprocessing in the accretion disc, it is a good tracer of the disc transfer function. In particular, the variability associated with the iron line can be used to track the evolution of the inner disc radius, as shown in \citet{RGC99}, \citet{gilfanov00} and \citet{RGC01}.
We repeat their approach here for the sake of completeness.
The time-dependence of flux of the reflected emission $F_{\rm refl}(E, t)$ (where $E$ is the photon energy) can be expressed through the convolution of the incident X-ray flux $F_{\rm inc}(E, t)$ with the disc transfer function $T(t)$:
\begin{equation}
    F_{\rm refl}(E, t) = \int\limits_{0}^{\infty} F_{\rm inc}(E, t-t') T(t') dt'.
\end{equation}
Following the convolution theorem, the Fourier image of reflected emission, $\hat{F}_{\rm refl}(E, f)$, is the pointwise product of the Fourier images of the incident emission, $\hat{F}_{\rm inc}(E, f)$, and the transfer function, $\hat{T}(f)$:
\begin{equation}
    \hat{F}_{\rm refl}(E, f) = \hat{F}_{\rm inc}(E, f) \hat{T}(f).
\end{equation}
Using the total frequency-resolved spectra $\hat{F}_{\rm tot}(E, f)=\hat{F}_{\rm refl}(E, f)+\hat{F}_{\rm inc}(E, f)$, we compute the (frequency-dependent) EW as
\begin{equation}
  EW(f) = \sum\limits_{E} \left|1-\frac{\hat{F}_{\rm tot}(E, f)}{\hat{F}_{\rm inc}(E, f)}\right| \Delta E \propto \frac{|\hat{F}_{\rm refl}(E, f)|}{|\hat{F}_{\rm inc}(E, f)|} = |\hat{T}(f)|,
\end{equation}
where the summation is over the energy bins.
Hence, the EW of the iron line can be used as a direct probe of the disc transfer function.

In our analysis we follow the same procedure as described in \citet{RGC99}. Light curves are extracted from 3\,keV to 9.6\,keV in bins of width $\Delta E=0.3$\,keV. Above 7.2\,keV the count rate becomes quite low, forcing us to use bins of 0.6\,keV. The final bin is extends from 8.4 to 9.6\,keV. A power spectrum is calculated for each energy range, and integrated to determine the contribution from each frequency range chosen: 0.01$-$0.02, 0.02$-$0.04, 0.04$-$0.08, 0.08$-$0.2, 0.2$-$0.4, 0.4$-$0.8, 0.8$-$1.6, 1.6$-$3.2, 3.2$-$6, 6$-$9 and 9$-$12\,Hz. We also evaluated the coherence (using 2--10 keV as the reference band), and confirmed that it is high ($\geq$95\%) in the energy and frequency ranges used. This allows us to create a spectrum corresponding to each frequency range for Epoch 1 to 3. In the later epochs, the combination of weaker variability and lower count rates at higher energies mean that we are unable to trace the iron line even with a reduced resolution of 0.3\,keV. Fig.~\ref{fig:ewfreq_fits} shows examples of the frequency-resolved spectra obtained.

Each spectrum was fit with a model (solid lines in Fig.~\ref{fig:ewfreq_fits}) comprising a power law for the incident continuum and a gaussian component for the line (the energy resolution of 0.3\,keV means that this is a reasonable approximation). The width of the gaussian was allowed to vary between 0.1 and 1\,keV. The best-fit index of the power law was close to that of the time-averaged spectra, but for Epoch 1 the index in the frequency-resolved fits was somewhat (a shift of $\sim0.1$) softer. After a good fit was found, the EW of the iron line was calculated and the uncertainty was derived from the fit parameters of the gaussian component.

Reprocessing of the X-rays may also take place in the cool outflowing material -- the disc wind, which spectroscopic and polarization signatures have been detected during the hard state \citep{MunozDarias19,kosenkov20}.
This additional component may affect the flux at the core of the line and its variability.
We therefore also tried to model the line using two gaussian components, one broad and one narrow. However, the data do not allow both components to be constrained. With the diminished resolution required to provide an adequate signal-to-noise ratio, a single gaussian is able to provide a good fit to the line, and this simple model is thereby sufficient for our analysis.

Fig.~\ref{fig:ewfreq_newresp} shows the resulting evolution of the EW as a function of frequency throughout Epochs~1 to 3.
Although the uncertainties are large, we can see significant changes between the observations. In Epoch~1 (left panel), early in the outburst, the iron line cannot be detected at all above 0.1\,Hz, and we can only determine the upper limits. The EW of the line at lower frequencies is below 100~eV. In Epoch~2 (middle panel), the line can be detected at frequencies up to $\sim5$\,Hz. The EW is also greater, reaching up to 300~eV. In Epoch~3 (right panel), there is no detectable cut off, although the data do not allow us to go higher than 8~Hz as the count rate becomes too low ($\sim10$\,cts~s$^{-1}$) in the highest energy bins. 
The values of the EW here are again around 150\,eV. 
The absolute values of EW of the iron line in MAXI~J1820+070 are thus similar to those seen in the hard state of both Cygnus~X-1 and GX\,339--4 \citep{RGC99,RGC01}. 

The constancy of the EW at low frequencies suggests that the variations of the incident and reflected continuum have the same frequency dependence, while the suppression at high frequencies suggests that the amplitude of variations of the reflected emission becomes progressively smaller as compared to the variations of the incident X-ray continuum (note that there is no suppression of the incident continuum variability, as seen in panels B of Fig.~\ref{fig:comparison}).
We note that there is a dip in the EW around 0.1\,Hz, matching the frequency of the quasi-periodic oscillation seen in the power spectrum (although the large uncertainties point to a low significance for this feature).

The suppression of the EW at higher frequencies indicate that the characteristic size of the region where the line forms becomes smaller as the source moves to softer spectral states during the outburst. The highest frequency of the ``plateau'' changes from $\lesssim$0.015~Hz in Epoch~1 to $\gtrsim4$~Hz in Epoch~3, indicating that the size scale of the reflector changes by a factor of 250.

\section{Discussion}

Much effort has been made in trying to determine the geometry of the inner accretion flow, in particular in the (bright) low/hard state. Different methods often find different, and not seldom conflicting, results. The first indications of disc truncation came from spectral studies. A change from a standard accretion disc to a geometrically thick optical flow was early suggested as an explanation for the spectral states \citep[e.g.,][]{Ichimaru77,Esin97,PKR97}. A truncated disc scenario also explains the observed correlation of the X-ray reflection strength with the width of the iron line \citep{GCR99,RGC01}, and further support comes from the evolution of the spectral shape during the outburst \citep[favouring the change of the source of seed photons for Comptonization;][]{kajava16} as well as the evolution of the short-term optical/X-ray correlation \citep{veledina17}. The picture is usually challenged by studies of the relativistically broadened iron line, which favour inner disc radii close to $\lesssim 1R_{\rm S}$ in the bright hard state and often require the black holes to be highly spinning \citep{fabian14,Garcia15}.

The accretion geometry can also be probed using temporal studies. The energy dependence of both variability and lags between spectral bands require spectral stratification such that rapid variability (produced in the inner regions of the disc) are connected to harder spectra \citep[e.g.,][]{MCP00,KCG01,PV14,AD18}. This is naturally achieved when the disc is truncated, and replaced by a hot inner flow at smaller radii. Such a geometry is also the basis for the model of the QPOs as Lense-Thirring precession of the hot inner flow \citep[][]{FB07,ID11}; this model can be used to predict the truncation radius and shows agreement with values inferred from the low-frequency cut-off of the power spectra. Studies of reverberation, where the reprocessing of the high-energy photons in the accretion disc introduces soft lag features in the energy-dependent high-frequency variability, often find lags pointing to disc truncation \citep[e.g.,][]{DeMarco15reverber}. However, there are also reverberation studies which find small lags, interpreted as a small inner disc radius in the bright low/hard state \citep{uttley14reverberation}. This disagreement between measured inner disc radii exists for the same source (\citealt{plant15,DeMarco15,basak16,basak17,demarco17}; see also fig.~11 in \citealt{Garcia15} for a compilation of different results); however, the models where the cold disc is not truncated in the hard state have difficulties explaining the QPO of the iron line centroid \citep{IvdK16}.

In the case of MAXI~J1820+070, a picture with a lamp-post and a non-truncated disc has been put forward to explain timing properties of the iron line and thermal reverberation lags \citep{Kara19}. 
The short time delays and a shift of the thermal reverberation lags towards higher frequencies found in (our) Epoch~2 onward prompted their interpretation of a small truncation radius and a compact and contracting X-ray corona. Moreover, the computed intrinsic lags of the iron line with respect to the continuum were found to be consistent with the thermal reverberation lags.
Another pillar of the contracting corona model comes from spectral modelling of the iron line region in \citet{Buisson19}, who advocate a stable inner radius of the disc during the hard and hard-intermediate states as evidenced by the stable shape of the red wing of the iron line.

The lamp-post picture for MAXI~J1820+070 was challenged in \citet{zdziarski21}, who find that the iron line shape can instead be described by reflection of two X-ray continua, one of them on the top of the inner parts of the truncated disc, and the other is within its truncation radius. Further, \citet{deMarco21} use the covariance function to extract the thermal component (associated with the reverberation lags) from the total spectra. 
They show that the temperature of this component significantly changes throughout the rising phase and state transition, in agreement with the scenario of the decreasing inner disc radius.
Here we discuss the application of the lamp-post and truncated disc geometries to the BH binary MAXI~J1820+070 in terms of the results of our analysis.

\subsection{Evolution of the broadband variability}

The overall evolution of both the total rms variability and power spectrum found in MAXI~J1820+070 (Figs.~\ref{fig:rmslinecomp}a and columns B and C of Fig.~\ref{fig:comparison}) matches that seen in other black hole transients, in particular GX\,339--4 \citep[e.g.,][]{BM16}.
The increase in rms towards lower energies (seen in Epoch~1) has previously been explained by the varying Comptonization continuum, which has a pivoting point around 50~keV \citep{GZ05}.
This interpretation faces difficulties with the observed difference of rms-energy dependence at different Fourier frequencies (Fig.~\ref{fig:comparison}, C1).
Alternative explanations have involved the cold accretion disc, which gives substantial contribution to the soft energy band and is intrinsically variable at frequencies $\lesssim0.1$~Hz \citep{uttley11,DeMarco15}. 
The disc emission has to have a variability amplitude higher than that of the Comptonization continuum in order to comply with the rms enhancement at lower energies \citep{Dzielak21}. 
This scenario is in line with the different behaviour of the rms at different frequencies; however, it faces difficulties in explaining the evolution of the rms between our Epochs~1 and 2. At later stages of the outburst, when the disc contribution to the soft energy band is stronger, we instead see a suppression at low energies, supporting earlier suggestions that the disc is stable at these frequencies \citep[e.g.,][]{churazov01,MP06}. It is difficult to explain how the disc can change variability behaviour between observations. Invoking the presence of a region in the disc irradiated by the X-ray source \citep[e.g.,][]{GDP09} does not solve the problem either, as in this case it is not clear what causes the drop of the rms at later stages (from Epoch 2 onwards), when the Comptonization continuum is clearly present, and the disc is expected to remain illuminated.

A promising mechanism producing the observed complex dependence of the rms on energy, Fourier frequency, and the evolution of these characteristics over the course of the outburst involves the presence of two Comptonization continua, one of which has a pivoting point in the {\nicer} energy range \citep{V18}. At low Fourier frequencies, we would then see variability from both components, while at high frequencies, the constant-slope component is suppressed, and the high rms above $\sim2$~keV is explained by the pivoting component.
In this scenario, we expect to see an upturn of the rms-energy spectra at high energies, above about 10~keV, similar to what has been observed in GX~339--4 with \textit{RXTE} data \citep{belloni11}.
Detailed modelling of the evolving aperiodic variability properties in such a framework will be presented in a separate work.

The presence of two types of variability, constant-slope and pivoting, has been observed in the X-ray binary Cyg~X-1 \citep{ZPP02}, and a requirement of two Comptonization continua in MAXI~J1820+070 was recently put forward by \citet{zdziarski21}, based on spectral analysis of the 2018 outburst using \textit{NuSTAR} data. They find a preferred geometry where the hard X-rays are produced in a central Comptonization region with large scale height, with a second soft Comptonization region placed at somewhat greater radii, at least partially overlapping the disc. In this scenario, it seems likely that it is the soft Comptonization region which produces the variability we see at low energies in Epoch 1, and which is suppressed as the stable disc moves to smaller radii during the outburst, replacing the contribution of the soft Comptonization at lower energies.

\begin{figure}
   \centering 
   \includegraphics[width=0.45\textwidth]{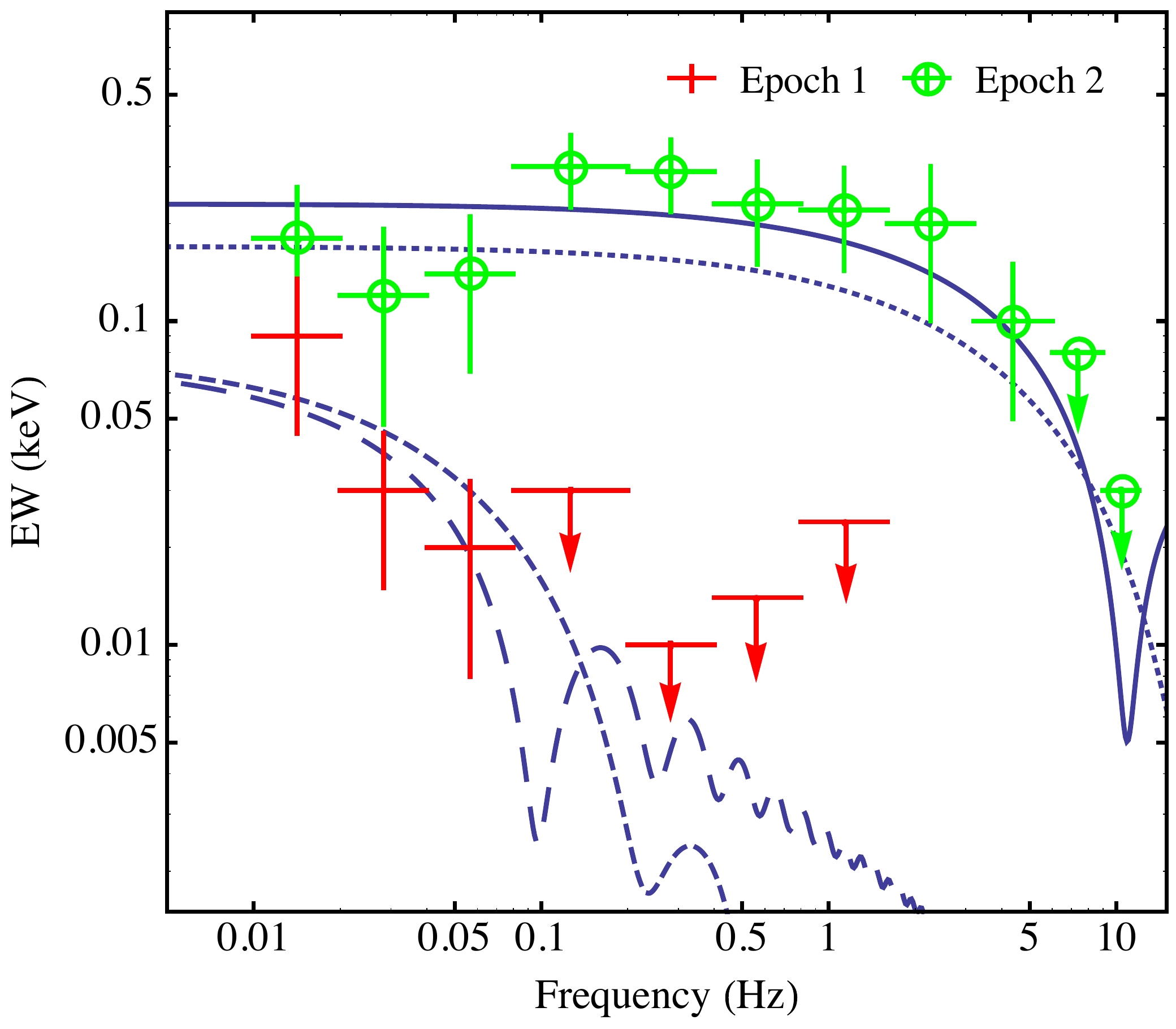}
     \caption{Qualitative modelling of EW as a function of Fourier frequency for Epochs 1 and 2 using two different models: a truncated disc and a lamp-post geometry. The inclination is set to $i=75\degr$, and the disc is assumed to have a constant scale height of $H/R=0.1$ and extend out to $10^6\,R_{\rm S}$. In both cases, the source is placed at $R=0$. For the truncated disc model, the source is placed at a height 0.3\,$R_{\rm S}$ and lines show the transfer function assuming the disc is truncated at an inner radius of $3\times10^3\,R_{\rm S}$ (long-dashed line) and $10^2\,R_{\rm S}$ (solid line). For the lamp-post geometry, the disc extends inward to $R_{\rm in}=0.618R_{\rm S}$, and the source is placed at a height of $3\times10^3\,R_{\rm S}$ (dashed line) or $10^2\,R_{\rm S}$ (dotted line) above the disc plane.
    } 
\label{fig:ewmodels1}
\end{figure}

\subsection{Evolution of the iron line profile}

Although detailed spectral modelling of the iron line lies outside the scope of this study, it is clear that the iron line region changes throughout the outburst. 
Interestingly, Fig.~\ref{fig:rmslinecomp}b 
seems to indicate that the changes are mainly in the core and blue wing of the line -- the red wing seems quite stable by comparison. The same results were found by \citet{Kara19} and \citet{Buisson19}. There could be many reasons for such variations, e.g., changes in ionization or scattering in hot/outflowing material \citep[e.g.,][]{garcia10}. For example, increasing ionization of the wind towards the soft state was indicated by the disappearance of the P-Cyg line profiles in the optical range \citep{MunozDarias19}, while the presence of winds themselves were still evidenced by the lines in the near-infrared range \citep{sanchez-sierras20}.

Furthermore, the inclination of the system will impact the peak energy of the iron line, with the strongest effect seen for small disc radii \citep{laor91}. This coupling between a broad shape of the iron line and inclination is a decisive factor also in the results of \citet{Buisson19}, who find an inclination of $\approx 34\degr$, drastically different from the measured angle of the orbital \citep[$\sim$$75\degr$,][]{Torres19} and jet \citep[$\sim$$63\degr$,][]{atri20,espinasse20} inclinations. This low value for the inclination, inferred from the iron line, is needed for the peak of the iron line to lie at the observed energy of 6.5\,keV.
The peak of the line produced at moderate disc radii, between about $3$ and $20R_{\rm S}$, will be blueshifted due to the dominant role of special relativistic effects (Doppler shift). While this effect is minor at inclinations $i\lesssim30\degr$, it becomes increasingly important for high inclinations: at inclinations higher than about $55\degr$, the overall line peak will be located at $7$~keV \citep{dauser10,wilkinsfabian11,poutanen20}.
On the other hand, at inclinations smaller than $30\degr$ the line Doppler shifts are negligible, and we instead expect the gravitational redshift effect to completely dominate, reducing the iron line peak to energies below $6.4$~keV.
The interplay between the Doppler shift and gravitational redshift effects is optimal at $\sim30\degr$ to obtain the peak of the line at energy close to that of the rest frame. 

The relatively small changes in the iron line seen during the evolution of the hard state observations were used to support a stable inner disc radius at the ISCO \citep{Kara19,Buisson19}. However, the same argument would mean that we expect the disc to be at the ISCO also in our Epoch 1, very early on in the outburst. As noted above, this makes it hard to explain the differences in variability for Epoch 1. Instead, it seems more likely that the changes seen in the line profile during Epochs~4 and 5, after the observations considered in \citet{Kara19} and \citet{Buisson19}, are connected to a decreased inner disc radius. The line peak then appears at about 7~keV, consistent with the expectation of the maximal energy for an inclined disc. In principle, emission from H-like iron (Fe~XXVI) could be contributing at this energy, but we do not expect to see it increasing in prominence with the observed decrease of hard X-rays (spectral softening). Rather than trying to find reasons for enhancing the emission at $\sim$7~keV, we find it more likely that the shift in line peak signifies that the disc reaches the ISCO in Epochs 4 and 5. This is also where the covariance spectra show that the disc inner radius stabilizes, although there are signs that the disc truncation radius is again larger between Epochs 5 and 6 \citep[][]{deMarco21}.

\subsection{Evolution of the iron line timing properties}

Our analysis showed a possible suppression of the rms around the iron line energies in Epochs 1--5 (Fig.~\ref{fig:rmslinecomp}a). Such a suppression has previously been observed in GS~2023+338 \citep{Oosterbroek96} and was  interpreted as signs of the line being formed far out in the reflector, possibly the outer accretion disc. In our case, this suppression is not significant, and we can access the characteristic radius of the reflector more carefully using our results from frequency-resolved spectroscopy.

\begin{figure}
  \centering 
  \includegraphics[width=0.45\textwidth]{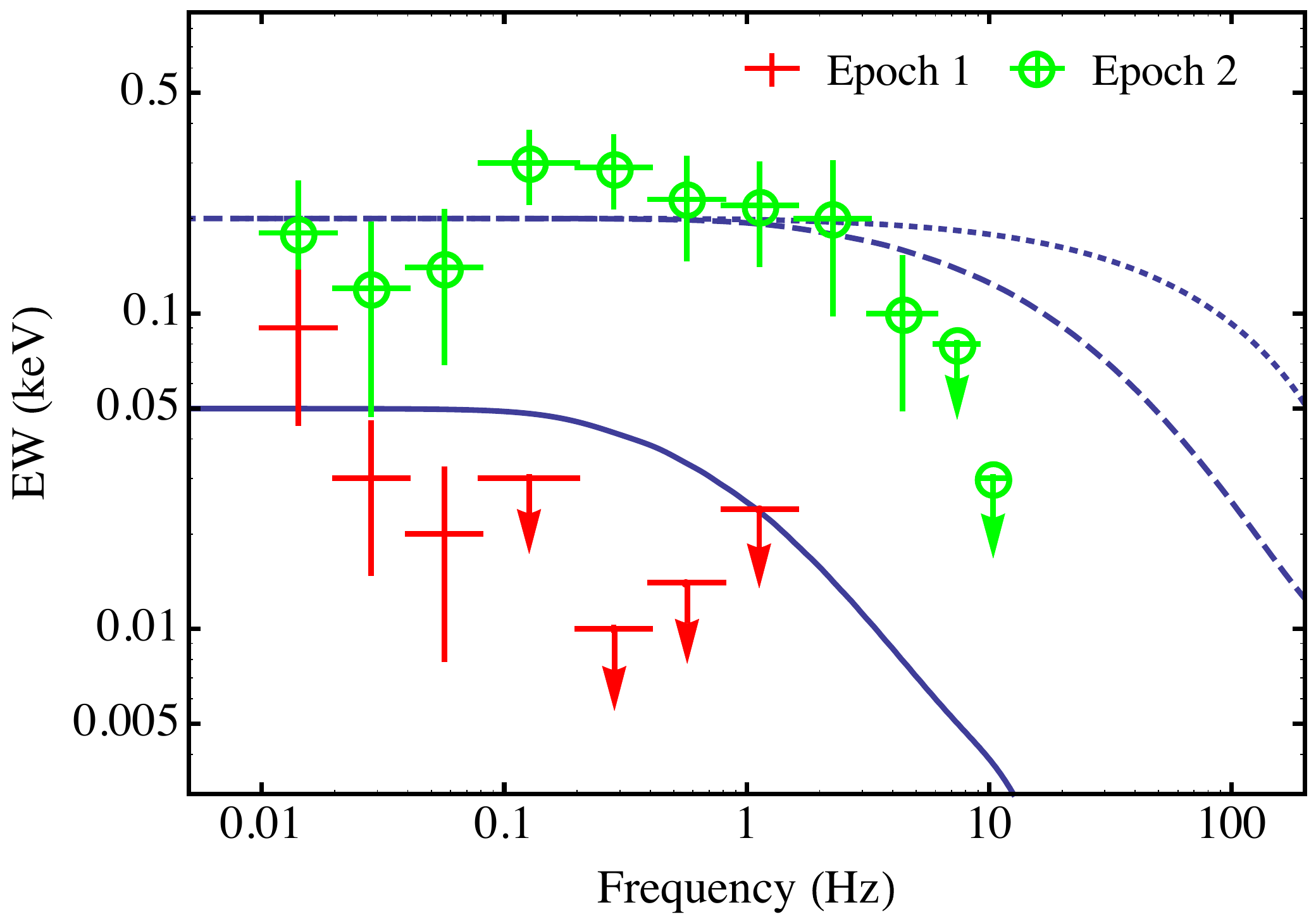}
     \caption{
     Models of the EW dependence on the Fourier frequency.
     The lines show the transfer function for the lamp-post geometry assuming the source is placed in the disc axis with $R_{\rm in}=0.617 R_{\rm S}$ with a height above the disc plane of $H_{\rm X}=2\,R_{\rm S}$ and inclination $i=34\degr$ (dotted line), $15\,R_{\rm S}$ and $i=45\degr$ (dashed line) for our Epoch~2, and $H_{\rm X}=50\,R_{\rm S}$ and $i=34\degr$ (solid line) for Epoch~1.
     The parameters are taken from \citet{Kara19} and \citet{Buisson19}.
      } 
\label{fig:modelcomparison3}
\end{figure}

We first try to model the disc transfer function for both the lamp-post and truncated disc geometries to see what parameters are needed to match our results. We stress that this is a rough estimate, not a fit to the data.
For the truncated disc geometry, we compute the response function of a disc with a central hole following the formalism described in \citet{Pou02}.
We assume the BH mass is $M=7M_{\odot}$ (where $M_{\odot}$ is the solar mass) and the system inclination is $i=75\degr$ \citep{Torres19}.
To mimic the truncated disc geometry, we assume the X-ray source is located close to the disc plane, at height $H_{\rm X}=1R_{\rm S}$.
The disc is assumed to have aspect ratio $H/R=0.1$, extending from (varying) $R_{\rm in}$ to $R_{\rm out}=10^6 R_{\rm S}$ (fixed) and be non-flared.
A flared disc for this source has been suggested in \citet{zdziarski21}. The difference between the flared and non-flared disc can be seen at low frequencies, and mostly affects phase/time lags, with minor modification to the shape of the response function \citep{Pou02}, justifying our choice of the non-flared disc.
For the lamp-post, we assume a disc inner radius $R_{\rm in}=0.618R_{\rm S}$ (equal to $1.235$ gravitational radii, i.e., ISCO for a maximally spinning Kerr black hole) and vary the height of the X-ray source $H_{\rm X}$. The other parameters are assumed to be the same as in the truncated disc geometry. Fig.~\ref{fig:ewmodels1} shows the model curves overlaid with the data.

To reproduce the phase lags, \citet{Kara19} use $i=45\degr$, $M_{\rm BH}=10M_{\odot}$, $R_{\rm in}=0.617R_{\rm S}$, $R_{\rm out}=500R_{\rm S}$ and find $H_{\rm X}\sim15R_{\rm S}$.
\citet{Buisson19} do spectral modelling using two lamp-post X-ray sources: one located at $H_{\rm X,inner} = 2R_{\rm S}$ and the other shifting from $H_{\rm X,outer}\sim50R_{\rm S}$ to $\sim 20R_{\rm S}$ for the times of our Epoch~1 and 2, respectively.
They additionally find $R_{\rm in}=2.65R_{\rm S}$ and $i=34\degr$.
We compare the transfer functions for $H_{\rm X} = 2$, 15 and 50 $R_{\rm S}$ in Fig.~\ref{fig:modelcomparison3} with our results of the EW for Epochs 1 and 2, where we could determine a cut-off, by scaling the low-frequency (constant) part of the model to the data.
As seen in the figure, all three cases severely overestimate the characteristic damping frequency.

Summarizing our results from the EW, the data require a very large characteristic distance between the X-ray source and the reflector in both geometries. In the lamp-post geometry, this means that the X-ray source needs to be placed far above the disc, at least $\sim3\times10^3R_{\rm S}$. It is difficult to explain the presence of a luminous ($\geq 10^{36}$\,erg~s$^{-1}$ using the \nicer\, flux) X-ray source at this distance. Both \citet{Kara19} and \citet{Buisson19} propose a geometry with a vertically extended corona, where the upper region is connected to reflection in the outer parts of the disc and thereby connected to the core of the iron line. However, the height found in \citet{Buisson19} is around $50\,R_{\rm S}$, well below the values required in Fig.~\ref{fig:ewmodels1}.

On the other hand, a truncated-disc geometry can produce significant EW of the iron line even when the inner radius is large, as long as the disc is flared, allowing a large fraction of the high-energy photons to be intercepted. Such a geometry for MAXI~J1820+070 has recently been suggested by \citet{zdziarski21}. They find a very low level of ionization in the reflector for the hard-state emission, indicating that the central X-emitting region is reflected off material quite far out in the disc (several hundred $R_{\rm S}$), well beyond the truncation radius. In order for the reflection fraction to be sufficiently high, the outer disc is assumed to be flared. This scenario fits well with the results presented here, and give a possible explanation as to why the size scale we find is much greater than the assumed truncation radius ($\sim10-50$ Schwarzschild radii). Placing the reflection at large radii also explains the relatively small changes to the iron line profile in the hard state, as the line profile is then not primarily driven by the location of the truncation radius. 

Moreover, \citet{Dzielak21} suggest a stratified structure of the hot flow with more rapid variability seeing hotter seed photons, consistent with propagating fluctuations. The same result would also be seen if the hardest component has a pivoting point at low energies, as suggested above. Both interpretations can be accommodated in the truncated disc scenario \citep{KCG01,V18}, but is more difficult to explain in a lamp-post geometry. In the latter case we would expect the soft photons to lag the hard ones at low frequencies, in contrast to what is observed.

As noted by \citet{Kara19}, the observational characteristics of MAXI~J1820+070 in many ways resemble those of GX~339--4, where an order of magnitude larger lags have been used to suggest disc truncation \citep{demarco17}. The smaller lags in MAXI~J1820+070, instead, seemingly signify a non-truncated disc. However, reverberation lags are strongly dependent on inclination, and smaller lags in MAXI~J1820+070 may be expected from its higher inclination (as compared to GX~339--4), rather than a physical difference in $R_{\rm in}$.
Namely, for the lamp-post model and non-truncated disc, one expects the reverberation lags of the order $\Delta t_{\rm lamp}\sim (1+\cos i) H_{\rm X}/c$, while for the truncated disc with an X-ray source in the middle the corresponding lag is $\Delta t_{\rm tr}\sim (1 - \sin i) R_{\rm in}/c$.
For the same physical distance from the X-ray source to the reflector, $H_{\rm X}=R_{\rm in}$, and for the inclination $i\gtrsim60\degr$, the reverberation lags in the truncated disc geometry are one-two orders of magnitude smaller than those of the lamp-post model.

Regardless of geometry, our results point to a large change in characteristic distance to the reflector, and that it decreases as the source moves toward softer states; however, we find that the disc is not at the ISCO during the bright(est) hard state (Epoch~2). For Epoch~3, the fact that we see no cut-off of the EW in the frequency range studied gives an upper limit of $\sim10\,R_S$, two orders of magnitude smaller than in Epoch~1.

\section{Conclusions}

We performed spectral and timing analysis of the evolving X-ray spectra of BH XRB MAXI~J1820+070 during the rising phase of the outburst.
Our main findings can be summarized as follows:
\begin{itemize}
    \item We observe change of the rms-energy dependence (Fig.~\ref{fig:rmslinecomp}a). In Epoch~1 the rms decreases with energy, while starting from Epoch~2 the rms is an increasing function of energy. We find a correlated change in the power spectra, where the high energies show greater variability at low frequencies in Epoch~1, and low energies dominate in the remaining epochs.
    \item We see no change of the shape of the iron line up to (and including) Epoch~3 (Fig.~\ref{fig:rmslinecomp}b). Starting from Epoch~4 we start to see excess emission above $\sim$7~keV, and the peak of the line shifts from 6.5 to 7~keV.
    \item The characteristic variability frequencies of the iron line, as probed by frequency-resolved spectroscopy, increases from Epoch 1 onward (Fig.~\ref{fig:ewfreq_newresp}). Between Epochs 1 and 3, the damping frequency increases by at least two orders of magnitude.
\end{itemize}

We show that the geometry with the lamp-post X-ray source and a stable disc inner radius have severe difficulties in explaining the shape and variability of the iron line: physical explanation for the shift of the iron line peak remains elusive, and the dependence of the EW on the Fourier frequency cannot be reproduced for the lamp-post heights deduced from the previously-found reverberation lags or spectral fits.
On the other hand, the truncated disc-inner hot flow geometry is able to accommodate our results. 
The rms-energy dependence and its evolution with time may be attributed to the presence of two Comptonization continua, which contribution is being replaced, at soft energies, by the more stable, cool accretion disc.
A large truncation radius together with large inclination will give a small damping frequency, as evidenced by the EW, and the small phase lags previously reported for this source.
As the disc truncation moves to smaller radii, it dampens the low-energy rms. 
Furthermore, as the optically thin accretion flow in the vicinity of the black hole is gradually replaced by the cool, optically thick accretion disc, we expect an appearance of the relativistically broadened iron line. 
For the source at high inclination, such as MAXI~J1820+070, its peak should be located at $\sim7$~keV, in agreement with the one observed on/after Epoch~4.
Our modelling suggests that the disc inner radius decreases both during the hard state and state transition. 
We derive the truncation radius of the order of $R_{\rm in}\sim10^3R_{\rm S}$ in the rising hard state (Epoch~1), which then decreases to $\sim100R_{\rm S}$ in the brightest hard state (Epoch 2) and reaches $\lesssim10R_{\rm S}$ at the hard-to-soft-state transition (Epoch 3).

\section*{Acknowledgements}

This research has made use of data and software provided by the High Energy Astrophysics Science Archive Research Center (HEASARC), which is a service of the Astrophysics Science Division at NASA/GSFC.
We thank Juri Poutanen for sharing the code for the disc transfer function and for the comments on the paper.
AV acknowledges the Academy of Finland grant 309308.
We acknowledge support from the International Space Science Institute (Bern).

\section*{Data availability}

The  data  underlying  this  article  are  available in  HEASARC, at \url{https://heasarc.gsfc.nasa.gov/docs/archive.html}.




\bibliographystyle{mnras}
\bibliography{allbib}








\bsp	
\label{lastpage}
\end{document}